\documentclass[useAMS,usenatbib]{mnras}

\usepackage{graphicx}   
\usepackage{amsmath}    
\usepackage{amssymb}    
\usepackage{multicol}   
\usepackage{bm}         
\usepackage{pdflscape}  
\usepackage{float}      
\usepackage{tipa}       
\usepackage{amsmath}
\usepackage{breqn}
\usepackage[]{enumitem}
\usepackage{color, soul}

\hypersetup{final}

\title[]{Hot Jupiter accretion: 3D MHD simulations of star-planet wind interaction}
\author[]{Daley-Yates S., Stevens I. R.}

\author[S. Daley-Yates, I. R. Stevens]{S. Daley-Yates$^{1}$\thanks{E-mail:
sdaley@star.sr.bham.ac.uk}, I. R. Stevens$^{1}$\\
$^{1}$School of Physics and Astronomy, University of Birmingham, Edgbaston, Birmingham, B15 2TT\\}

\begin{document}

\date{}

\pagerange{\pageref{firstpage}--\pageref{lastpage}} \pubyear{2018}

\maketitle

\label{firstpage}

\begin{abstract}
We present 3D MHD simulations of the wind-wind interactions between a solar type star and a short period hot Jupiter exoplanet. This is the first such simulation in which the stellar surface evolution is studied in detail. In our simulations, a planetary outflow, based on models of FUV evaporation of the exoplanets upper atmosphere, results in the build-up of circumstellar and circumplanetary material which accretes onto the stellar surface in a form of coronal rain, in which the rain is HJ wind material falling onto the stellar surface. We have conducted a suite of mixed geometry high resolution simulations which characterise the behaviour of interacting stellar and planetary wind material for a representative HJ hosting system. Our results show that magnetic topology plays a central role in forming accretion streams between the star and HJ and that the nature of the accretion is variable both in location and in rate, with the final accretion point occurring at $\phi~=~227^{\circ}$ ahead of the sub-planetary point and $\theta~=~53^{\circ}$ below the orbital plain. The size of the accretion spot itself has been found to vary with a period of $67 \ \mathrm{ks}$. Within the accretion spot, there is a small decrease in temperature accompanied by an increase in density compared with ambient surface conditions. We also demonstrate that magnetic fields cannot be ignored as accretion is highly dependent upon the magnetic topology of both the HJ and the host. We characterise this behaviour as Star Planet Wind Interaction (SPWI).
\end{abstract}


\section{Introduction}
\label{ACCRETIONsec:intro}

Hot Jupiter (HJ) type exoplanets reside in short orbits, having undergone migration from the orbital distances at which they form, these planets are subject to intense radiation from their host stars \citep{Fogg2005, Murray-Clay2009, Petrovich2016, Alvarado-Montes2017}. As their radii and masses are similar to the solar system planet Jupiter, they are the largest and most readily detectable class of exoplanets. Indeed the first exoplanet to be discovered, 51 Peg b by \cite{Mayor1995} falls into this category. Due to the short periods of their orbits ($\sim 4 \ \mathrm{days}$), HJ are very suitable to detection via the transit and radial velocity methods, see \cite{Wright2012} for a comprehensive review of detection methods. Despite this observational bias, many questions about the nature of HJ systems still remain unanswered and unpredicted observations unexplained. An example is the so called HJ \textit{pile-up}, a phenomenon where the orbital period of HJs clusters between 2 - 6 Earth days and trails off to longer periods, giving rise to the term \textit{pile-up}, see \cite{Chang2012} for possible responsible mechanisms and more resent theoretical work by \cite{Wise2018, OConnor2018}. Such unexpected statistical features hint at the limit of our understanding and of the future challenges to theorists and observers who study HJ systems.

The migration of giant planets through to the HJ phase influences the evolution of planetary systems from initial planet formation to the final stages of the systems life. This stage, when a gas-giant becomes a HJ, is when the planet leaves its final mark on its host. Mass and angular momentum exchange as well as magnetic interactions between the HJ and its host impact on the long term evolution of both bodies, acting to either reduce or enhance the rate of spin down and orbital decay. Such effects have been investigated theoretically by \cite{Strugarek2015} who found that the torque experienced by HJs can vary over an order of magnitude depending on the magnetic field of both HJ and host star; reporting migration timescales of 100 Myr. Such results show that star-planet magnetic interaction (SPMI) must be considered if a rigorous theory of the final stages of giant planet migration is to be developed. 

Our aim with the present study is to further explore the paradigm of star-planet interaction by focusing on a HJ undergoing hydrodynamic mass-loss of its atmosphere and to place the results in the context of observable signatures left on the surface of it host star via accretion flows. 

\subsection{Planetary FUV evaporation}
\label{ACCRETIONsec:UV}

Mass-loss in HJs occurs predominately due to absorption of energy via incident (Far Ultraviolet) FUV and X-ray flux from the host star \citep{Lammer2003, Pillitteri2015}. Mass is then lost either through Jeans escape, which is a function of the planet's effective temperature, or by thermal expansion driven by the aforementioned incident radiation. For HJs it has been shown that Jeans escape vastly underestimates their mass-loss. Indeed, the first exoplanet discovered, 51 Peg b, has an effective temperature similar to that of Jupiter's upper atmosphere ($\sim 700 - 1000 \ \mathrm{K}$) \citep{Smith1990}. It is the FUV and X-ray heating of the upper exosphere that results in temperatures an order of magnitude higher \citep{Lammer2003} than the effective temperature. This enhancement of the exosphere temperature is directly responsible for the hydrodynamic atmospheric expansion seen in multiple HJs \citep{Lammer2003, Weber2017, Sairam2018}. Enhanced mass-loss is not the only a consequence of absorbed stellar radiation. Many HJs exhibit lower mass while simultaneously a larger radii with respect to Jupiter \citep{Tremblin2017} and hence a lower density, a result consistent with an increased internal energy. This increased energy via FUV absorption is though to contribute to atmospheric circulation, leading to a redistribution and possible equalisation of day- and night-side temperatures (see \cite{Schwartz2017, Dang2018, Zhang2018} for observations of temperature offsets from the sub-stellar point). FUV absorption possibly contributes to enhancement of the planetary dynamo and hence magnetic field. Values of the order $50 \ \mathrm{G}$ are predicted by \cite{Yadav2017}, an order of magnitude larger than Jupiter's magnetic field. However, such field strengths would be detectable due to MHz radio emission via the Electron Cyclotron Maser Instability (ECMI) \citep{Weber2017, Daley-Yates2017, Daley-Yates2018}.

The estimated mass-loss for HJs is typically in the region of $10^{9} \ \mathrm{g/s}~<~\dot{M}_{\mathrm{HJ}}~<~10^{12} \ \mathrm{g/s}$ \citep{Salz2016}. The dynamics and spread of the evaporated HJ atmosphere in the interplanetary medium behaves differently depending on the rate of mass-loss. At lower $\dot{M}_{\mathrm{HJ}}$, the evaporating HJ atmosphere fails to reach the stellar surface and is swept back forming a cometary tail \citep{Matsakos2015, Daley-Yates2018}. For comparatively large values of $\dot{M}_{\mathrm{HJ}}$, the atmosphere can undergo Roche lobe overflow and form strong accretion streams connecting with the stellar surface \citep{Matsakos2015, Pillitteri2015}. This form of interaction is the topic of the present work.

\subsection{Star planet wind interaction}
\label{ACCRETIONsec:windwind}

Recent observations of HJ hosting systems conducted with both the Cosmic Origins Spectrograph on board the Hubble Space Telescope (COS-HST) \citep{Pillitteri2015} and the Echelle Spectro Polarimetric Device for the Observation of Stars instrument at the Canada-France-Hawaii Telescope (ESPaDOnS-CFHT) \citep{Shkolnik2008} indicate that stellar emission is synchronised with the orbit of the HJ. It has been proposed that this is due to an intermittent accretion stream making foot-fall on the stellar surface, leading the HJ orbit by $\sim 90^{\circ}$. These two studies have also reported enhanced chromospheric activity for a number of stars hosting hot Jupiters, including HD 179949 and HD 189733 \citep{Shkolnik2008, Pillitteri2015}. Star-planet interaction via mass transfer, from planet to star, and thus the accretion of the planet by its host, is thought to be responsible. Thus far, there has been limited simulation work investigating this wind mediated type of star-planet accretion. Star-Planet Interaction (SPI) and Star-Planet Magnetic Interaction (SPMI) has been studied using Hydrodynamics (HD) and Magnetohydrodynamics (MHD) for a number of years (see below). 

Physical systems in which multiple accelerated gaseous outflows form collisions are numerous in astrophysics; from AGN jets to massive star colliding wind binaries. These interactions are responsible for a range of detectable emission from synchrotron and X-ray  to the acceleration of cosmic rays. In the context of exoplanetary systems and specifically those hosting HJs, there are two types of interaction depending on whether the HJ orbits inside or outside the stellar Alfv\'{e}n surface. Inside, the HJ can interact via direct magnetic field line connection between the stellar and planetary magnetospheres, this situation has been modelled by \cite{Strugarek2014a, Khodachenko2015, Vidotto2015, Strugarek2015a, Strugarek2016}. Outside the Alfv\'{e}n surface direct magnetic connection of field lines is not possible as the flow is super-Alfv\'{e}nic. Instead the mode of interaction is via magnetised flow of material from the HJ atmosphere to the stellar surface via an accretion stream (provided the HJ is undergoing rapid mass-loss), a type of interaction which will be referred to from here on as Star-Planet Wind Interaction (SPWI) and has been modelled previously by \cite{Bourrier2013, Bourrier2016, Owen2014, Alexander2015, Matsakos2015, Carroll-Nellenback2017}. Understanding the stellar wind dynamics is in itself an important aspect when considering SPI and there have been several separate studies that have investigated solely the properties of stellar winds of exoplanet hosting stars \citep{Alvarado-Gomez2016, Fares2017}. 

Another form of SPWI is the formation of planetary bow shocks around HJs which lead to observable effects such as ingress features in transit observations, where the bow shock obscures some of the stellar light before the main transit of the planet across the stellar disk \citep{Llama2013}. Bow shocks are features which form due to incident kinetic and magnetic energy in the stellar wind on the planetary magnetosphere. This energy input has been proposed as a means to amplify the ECMI process and has been previously thought to result in detectable emission \citep{Stevens2005, Zarka2007}. This has however recently been brought into doubt by \cite{Weber2017, Daley-Yates2018}.

These two categories of interaction, SPMI and SPWI, are complimentary; one does not preclude the other. Indeed, SPMI can occur at the same time as SPWI, however SPMI cannot occur if the orbit is outside the Alfv\'{e}n surface, where SPWI can as it is, in principle, independent of the planets position relative to the Alfv\'{e}n surface. The prerequisite for SPWI is that the HJ atmosphere be undergoing hydrodynamic escape and a planetary wind has established, a situation independent of the stellar Alfv\'{e}n surface position. Hydrodynamic escape is however conditional upon the absorption of stellar FUV radiation in the HJ upper atmosphere and is different for each HJ system. As a rule of thumb, a hydrodynamic wind will establish if the HJ orbital radius is $a~<~0.5 \ \mathrm{au}$, according to \cite{Weber2017}. The precise wind properties, mass-loss and day-side night-side outflows and temperature are a function of the stellar, planetary and orbital parameters, as briefly discussed in the previous section.

The focus of the present study is the quantification of the evolution of the stellar surface of a star which is undergoing accretion, via SPWI interaction, from the expanding atmosphere of a hosted HJ. Exploring non-magnetised and dipole-dipole interactions, we aim to distinguish between different regimes of magnetised interaction and quantify the mass accretion rate, accretion stream stability and angular momentum transfer between planet and host star.

\section{Modelling}
\label{ACCRETIONsec:modelling}

The models used here are derived from the work of \cite{Matsakos2015} and are identical to those we present in \cite{Daley-Yates2018}, with the exception of the planetary parameters for temperature and mass-loss (see Table \ref{ACCRETIONtab:params}). The reader is directed to this paper for a full description of the model used. Below, we give a short overview of the key equations.

\subsection{Magnetohydrodynamics}
\label{ACCRETIONsec:MHD}

To account for the orbital motion of the planet, the MHD equations are solved in the co-rotating frame of the planet and are given by:
\begin{equation}
\label{ACCRETIONeq:mass}
\frac{\partial \rho}{\partial t} + \bm{\nabla} \cdot \left( \rho \bm{v} \right) = 0
\end{equation}
\begin{equation}
\label{ACCRETIONeq:momentum}
\frac{\partial \bm{v}}{\partial t} + \left( \bm{v} \cdot \bm{\nabla} \right) \bm{v}  
+ \frac{1}{4 \pi \rho} \bm{B} \times  \left( \nabla \times \bm{B} \right) 
+ \frac{1}{\rho} \nabla p = \bm{g} + \bm{F}_{\mathrm{co}}
\end{equation}
\begin{equation}
\label{ACCRETIONeq:energy}
\frac{\partial p}{\partial t} + \bm{v} \cdot \bm{\nabla} p  
+ \gamma p \nabla \cdot \bm{v} = 0
\end{equation}
\begin{equation}
\label{ACCRETIONeq:magnetic}
\frac{\partial \bm{B}}{\partial t} 
+ \nabla \times \left( \bm{B} \times \bm{v} \right) = 0.
\end{equation}
Where $\rho$, $\bm{v}$, $\bm{B}$, $p$, $\bm{g}$ and $\bm{F}_{\mathrm{co}}$ are, density, velocity, magnetic field, pressure, gravitational acceleration and Coriolis and centrifugal acceleration. $\bm{F}_{\mathrm{co}}$ is the sum of the centrifugal and Coriolis forces: $\bm{F}_{\mathrm{co}}~=~\bm{F}_{\mathrm{centrifugal}}~+~\bm{F}_{\mathrm{coriolus}}$ and are:
\begin{equation}
\bm{F}_{\mathrm{centrifugal}} = 
- \left[ \bm{\Omega}_{\mathrm{fr}} \times \left( \bm{\Omega}_{\mathrm{fr}} 
\times \bm{r} \right) \right] = \bm{\Omega}_{\mathrm{fr}}^{2} 
\left( x \hat{x} + y \hat{y} \right)
\end{equation}
and
\begin{equation}
\bm{F}_{\mathrm{coriolus}} = 
- 2 \left( \bm{\Omega}_{\mathrm{fr}} \times \bm{v} \right) 
= 2 \Omega_{\mathrm{fr}} \left( x \hat{x} + y \hat{y} \right),
\end{equation}
where $\bm{\Omega}_{\mathrm{fr}}$ is the angular frequency of the rotating frame and $\bm{r}$ is the radial distance.

To close the MHD equations, we employ an adiabatic equation of state. To mimic the isothermal nature of the stellar and planetary winds, we set $\gamma~=~1.05$.

\subsection{Stellar and planetary models}
\label{ACCRETIONsec:models}

The initial conditions for the stellar wind and planetary mass-loss are based on the non-magnetic Parker wind model \citep{Parker1958}, the governing equation of which is
\begin{equation}
\label{ACCRETIONeq:parker_eq}
\psi - \ln{\left( \psi \right)} = - 3 - 4 \ln{ \left( \frac{\lambda}{2} \right)} 
+ 4 \ln{ \left( \xi \right) } + 2 \frac{ \lambda}{\xi}
\end{equation}
with $\psi$, $\lambda$ and $\xi$ being three dimensionless parameters which are defined as:
\begin{equation}
\psi \equiv \left( \frac{v^{\mathrm{init}}_{\mathrm{W}} (r) }{c_{\mathrm{s}}} \right)^2
\end{equation}
\begin{equation}
\lambda \equiv \frac{1}{2} \left( \frac{v_{esc}}{c_{\mathrm{s}}} \right)^2
\end{equation}
\begin{equation}
\xi \equiv \frac{r}{R}.
\end{equation}
$v^{\mathrm{init}}_{\mathrm{W}}(r)$ is the radial wind velocity profile at $t = 0$. The escape velocity is given by $v_{esc}~=~\sqrt{2 G M/R}$ and $c_{\mathrm{s}}~=~\sqrt{2 k_{\mathrm{B}} T / m_{\mathrm{p}}}$ is the isothermal sound speed. $k_{\mathrm{B}}$ is the Boltzmann constant, $T$ the temperature and $m_{\mathrm{p}}$ the mass of the proton. $\xi$ is the radial distance from either the centre of the star or planet, in units of stellar or planetary radii and $R$ is the radius of either body. 

\subsubsection{Stellar and planetary surface parameters}
\label{ACCRETIONsec:params}

The parameters for our model star and planet are presented in Table \ref{ACCRETIONtab:params} and are directly based on those used by \cite{Matsakos2015}, which in turn are parameterised values based on 1D simulations conducted by \cite{Matt2008} for base density and pressure. These values are modified so that mass-loss rates and wind properties in the 3D simulations agree with the 1D models.

To investigate different magnetic regimes, we conducted several simulation, each with a different combination of magnetic fields
\begin{itemize}[itemindent=4em]
	\item[\textbf{S0P0}] Non-magnetised interaction: the simulation is conducted in the HD regime, both the star and the planet have no magnetisation.
	\item[\textbf{S2P1}] Dipole-dipole interaction: the magnetic fields of both bodies are dipolar and aligned with the rotational axis of the system. Two simulations of this model were run, one Cartesian and one spherical polar, see Section \ref{ACCRETIONsec:Num} for details.
	\item[\textbf{S2P0}] Dipole-non-magnetised interaction: a combination of the first two topologies, the stellar magnetic field is dipolar and the planet has no magnetisation.
\end{itemize}
These regimes cover three different types of interaction and together they are designed to determine whether the stellar or planetary magnetic fields influences the planetary accretion. The names of each model indicates the magnetisation of each body, \textbf{S} stands for the star and \textbf{P} stands for the planet. The number following the letters are the strengths of the magnetic field of each body. So \textbf{S2P0} states that the star has a dipole equatorial magnetic field strength of $2 \ \mathrm{G}$ and the planet is not magnetised.

\begin{table*}
	\centering
	\caption[Stellar and planetary parameters used to simulate star--planet accretion.]{Stellar and planetary parameters used in the simulations.}
	\begin{tabular}{ccccc}
		\hline
		Parameter & Symbol & Star & Planet \\
		\hline
		Mass & $M_{\ast, \circ}$ & $1 \ M_{\odot}$ & $0.5 \ M_{J}$ \\
		Radius & $R_{\ast, \circ}$ & $1 \ R_{\odot}$ & $1.5 \ R_{J}$ \\
		Temperature & $T_{\ast, \circ}$ & $10^{6} \ \mathrm{K}$ & $10^{4} \ \mathrm{K}$ \\
		Equatorial magnetic field strength & $B_{eq \ast, \circ}$ & $0 \ \mathrm{G}$, $2 \ \mathrm{G}$ & $0 \ \mathrm{G}$, $1\ \mathrm{G}$ \\
		Surface density & $\rho_{\ast, \circ}$ & $5 \times 10^{-15} \ g/cm^{3}$ & $7 \times 10^{-16} \ g/cm^{3}$ \\
		Orbital radius & $a$ & $-$ & $0.047 \ \mathrm{au}$ \\
		Orbital period & $p_{\mathrm{orb}}$ & $-$ & $3.7 \ \mathrm{days}$ \\
		Rotational period & $p_{\mathrm{rot} \ast, \circ}$ & $3.7 \ \mathrm{days}$ & $3.7 \ \mathrm{days}$ \\
		\hline
	\end{tabular}
	\label{ACCRETIONtab:params}
\end{table*}

\subsection{Numerical modelling}
\label{ACCRETIONsec:Num}

Using the public MHD code PLUTO (version 4.2) \citep{Mignone2007, Mignone2011}, the MHD equations (\ref{ACCRETIONeq:mass} - \ref{ACCRETIONeq:magnetic}), were solved numerically using a 2nd order accurate scheme with linear spatial reconstruction (Van Lear limiter), 2nd order Runga-Kutta time-stepping and the HLLD Riemann solver. The zero divergence condition of the magnetic field was provided by the GLM method of \cite{Dedner2002}, see \cite{Mignone2010, Mignone2010a} for the PLUTO version.

Two types of numerical grids were employed in this study, the first, Cartesian, was designed to capture the global evolution of the extended wind of both the star and planet. Using adaptive mesh refinement (AMR), the Cartesian grid preserves and increases resolution at all radii from the stellar surface and where fluid features require a finer grid. 

The second grid, spherical polar, was used in only one simulation for model \textbf{S2P1}. The aim was to study the behaviour of material as it approaches the close in stellar magnetosphere, where material interacts with stellar surface. To capture this behaviour a spherical grid is superior as the common \textit{staircasing} effect which Cartesian grids suffer from when representing smooth or curved surfaces at low resolution is avoided. The restrictive boundary conditions of the Cartesian grid (see Section \ref{ACCRETIONsec:BCs}) also prevents the planetary wind material from interacting with the stellar surface, a spherical grid does not have this restriction.

There were a number of methods available to us to avoid the \textit{staircasing} effect. These included the brute force method of simply increasing the number of AMR levels at the stellar surface until the surface was approximately smooth over the dynamic length scale of the simulation. This option was however prohibitively computationally expensive. Another is the Level-set method which allows for the computation of smooth surfaces on numerical grids \citep{Osher1988}. The Level-set method is not a native feature in PLUTO therefore using a spherical grid proved more time-efficient for overcoming the \textit{staircasing} issue. 

\begin{figure}
	\centering
	\includegraphics[width=0.39\textwidth,trim={5cm 0cm 4.3cm 0},clip]{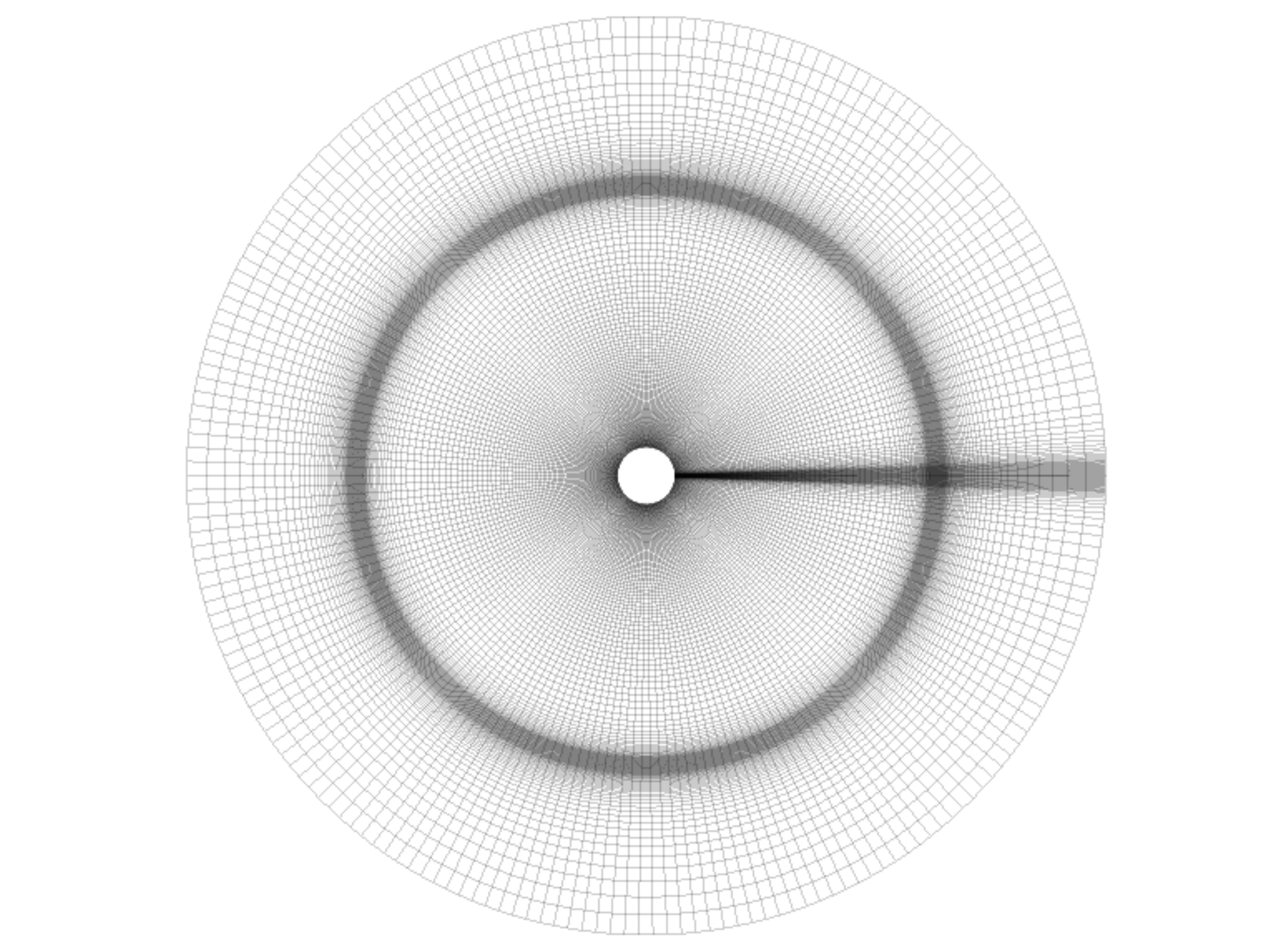}
	\includegraphics[width=0.39\textwidth,trim={5cm 0cm 4cm 0},clip]{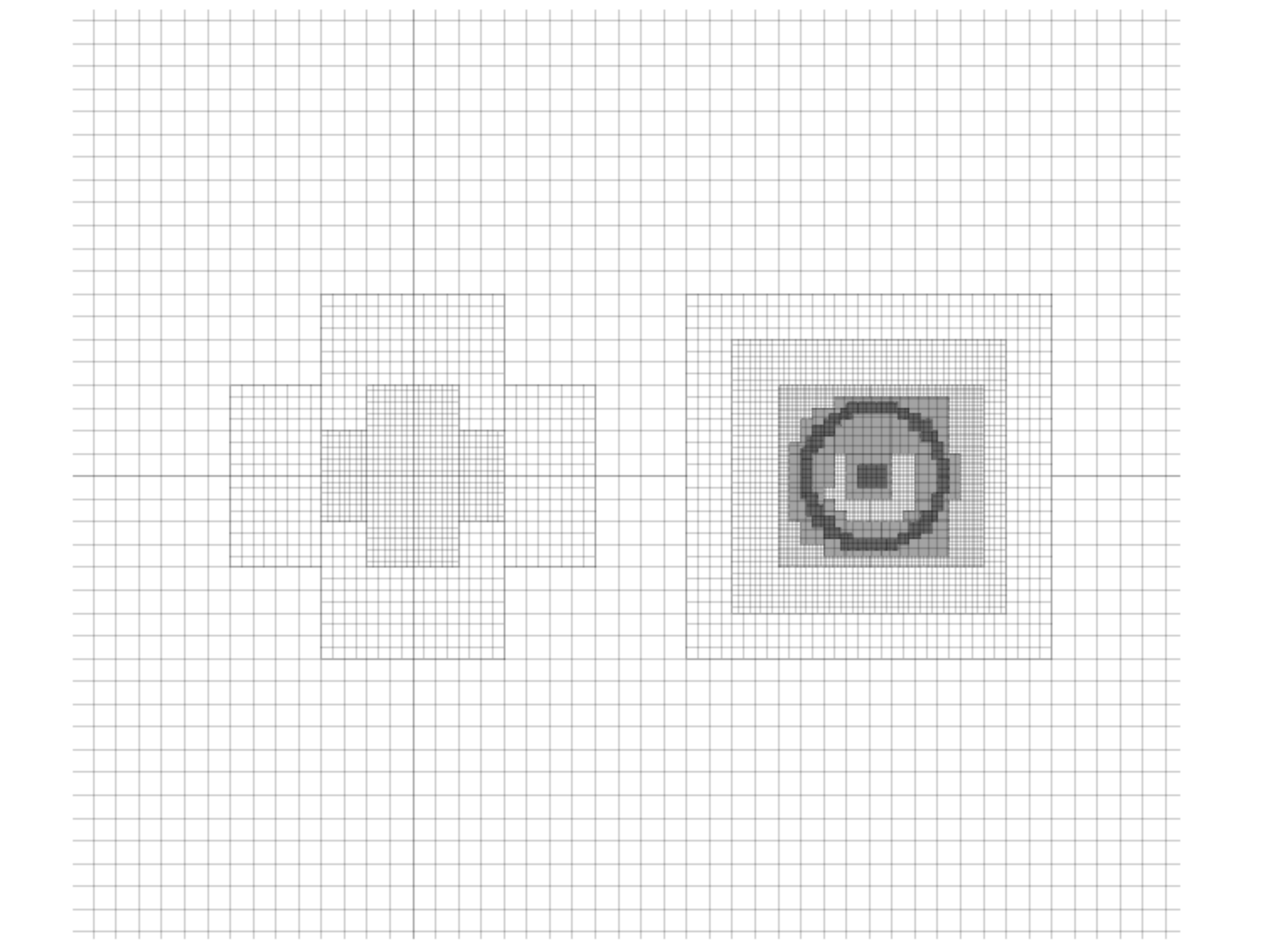}
	\caption[Examples of the grids used in the simulations of hot Jupiter accretion.]{Examples of the grids used in the simulations. Top: grid for the spherical polar simulation bands of darker lines indicate high resolution strips which coincide at the planet, allowing higher resolution where it is needed with minimal computational cost. Bottom: zoom in of the grid used for the Cartesian AMR simulations (the full grid is not shown here). The star is situated to the left and the planet to the right. The nested high resolution patches are clustered at points of high density gradient such as the stellar and planetary surfaces and the transition point between the stellar and planetary winds. The physical extent of both grids, as displayed here, are not to scale and are simply illustrative.}
	\label{ACCRETIONfig:grids}
\end{figure}

The following describes the two grid types, their extent and resolution. Examples of which are shown in Fig. \ref{ACCRETIONfig:grids}.

\subsubsection{Cartesian grid}

The Cartesian grid extended from $x, y~\in~\{ -32 \ R_{\odot}, 32 \ R_{\odot} \}$ and $z~\in~\{-16 \ R_{\odot}, 16 \ R_{\odot} \}$ with an initial resolution of $128^2 \times 64$ cells. This initial grid was then successively refined to a maximum of 5 AMR levels resulting in an effective resolution of $4096^2 \times 2048$ giving 64 cells/$R_{\ast}$. The initial grid setup is shown in the lower plot of Fig. \ref{ACCRETIONfig:grids}. Refinement of cells is determined by the gradient of the density across the cell, if the gradient exceeds a pre-specified amount, then the cell is marked for refinement. PLUTO uses the patch based method for the grid infrastructure and as such the regions marked for refinement have applied a patch of higher resolution cells covering a greater physical extent than the local cell which was marked for refinement.

\subsubsection{Spherical polar grid}

The spherical polar grid employed for model \textbf{S2P1} is defined by $r \in \{ 1\ R_{\odot}, 16 \ R_{\odot} \}$, $\theta \in \{ 0.02 \pi, 0.98 \pi \}$ and $\phi \in \{ 0, 2 \pi \}$ with a resolution of $178 \times 156 \times 252$. The grid at the $\theta$ boundaries is effectively clipped so that the grid does not touch the polar axis. This is done to avoid restrictive time-step sizes as the cell size diminishes at the poles, hence the range specified above.

As the planet is not at the centre of the coordinate system and due to the curvilinear nature of spherical polar grids, individual computational cells occupy different physical volumes on the day and night sides of the planet. High resolution is therefore required to avoid grid effects, as AMR is not used in this grid arrangement, the grid is stretched to ensure sufficient resolution at the planet to resolve the planetary outflow and minimise the difference in cell size between the day and night side. There are five distinct grid regions in all three spherical directions which are either stretched or uniform, allowing for the high resolution to be centred on the planet. An illustration of the stretched spherical polar grid is shown in the top image of Fig. \ref{ACCRETIONfig:grids}. In contrast to the Cartesian grid, the only region held constant in the spherical grid is within the planetary boundary, as the radial coordinate begins at the stellar surface, the star is effectively outside the grid and therefore not held constant.

\subsubsection{Boundary conditions}
\label{ACCRETIONsec:BCs}

The computational mesh is initialised everywhere according to the equations presented in Section \ref{ACCRETIONsec:models} for the stellar wind. A region of $10 \ R_{\circ}$ around the planet is initialised for the planetary wind in the same manner. 

In the Cartesian simulations, both the stellar and planetary bodies act as internal boundaries with stellar, planetary and wind parameters held constant out to 1.5 $R_{\ast}$ ($R_{\circ}$), see \cite{Daley-Yates2018} for more details. The outer boundaries are set to out-flowing insuring that edges of the computational domain do not influence the solution. This situation can change if the velocity begins to point inwards at the edge of the domain, resulting in the simulation being swamped by spurious material entering from the boundaries. This can be avoided by specifying that if $v_{r} < 0$ then override with $v_{r} = 0$. This condition was however not encountered in this study.

For the spherical polar grid, the inner radial boundary is set to the values given by equation (\ref{ACCRETIONeq:parker_eq}) for $\xi = 1$. The velocity here is held constant and positive to ensure inflow at the correct mass-loss rate for the star. The outer radial boundary is set to out-flowing. Both the upper and lower $\theta$ boundaries are set to reflective and the upper and lower $\phi$ boundaries are set to periodic to allow for material to flow between them. The planetary boundary conditions are identical to the Cartesian case, with a frame transformation and change of basis vectors providing the fluid fields in spherical coordinated form the Cartesian ones.

In both the Cartesian and spherical polar simulations, the outer magnetic field at the outer boundary of the system are set to outflowing.

\subsection{Tracking accretion}
\label{ACCRETIONsec:LatLongAcc}



To determine the exact position of material as it is deposited on the stellar surface, passive scalars are used to track the motion and advection of the planetary wind. Passive scalars exert no influence on the fluid dynamics but are advected by the fluid flow and can be used as a proxy for the motion of mass across the numerical grid. This allows for the position of the accreting quantities on the stellar surface to be defined.

At $t=0$, the passive scalars are set at the planetary surface and are continuously renewed as $t > 0$. As the simulation evolves, the passive scalars are advected into the computational active region. The initial planetary surface concentration of passive scalars is set to $1 \ n_{\mathrm{pass}}$ per unit volume (where $n_{pass}$ is the passive scalar concentration) and as they are advected, this concentration decreases or increases due to mixing of the stellar and planetary winds. The result is that the concentration of passive scalars can be used as a proxy for the density if normalized by the planetary surface density.


\section{Results and discussion}
\label{ACCRETIONsec:Res}

The following sections detail the results of the three models, describes the global behavior of the wind-wind interaction, the surface dynamics, accretion rates and finally places these results in the context of observable signatures. 

\subsection{Roche potential}
\label{ACCRETIONsec:Roche}

The physics of accretion occurs at all length scales in astrophysics from planetesimal formation on the shortest through massive star formation and black hole growth and Active Galactic Nuclei (AGN) at larger scales, form tori whose  jets influence the evolution of the host galaxy. While black hole and AGN accretion physics exist in the high energy accretion regime, stellar-HJ accretion occurs at much lower energy.

In the context of exoplanetary physics, accretion is most commonly explored when concerning the planets initial formation. In the present study we are concerned with accretion in the HJ phase of the planets life. For HJs, this means the flow, dynamics, physical quantities and transfer mechanics of photoevaporated atmospheric material from low mass object (planet) to high mass object (star). This is in contrast to stellar wind and stellar Roche lobe overflow accretion, where material is transferred from the physically larger object to the smaller, for example black hole or neutron star in the case of X-ray binaries.

\begin{figure*}
	\centering
	\includegraphics[width=0.49\textwidth]{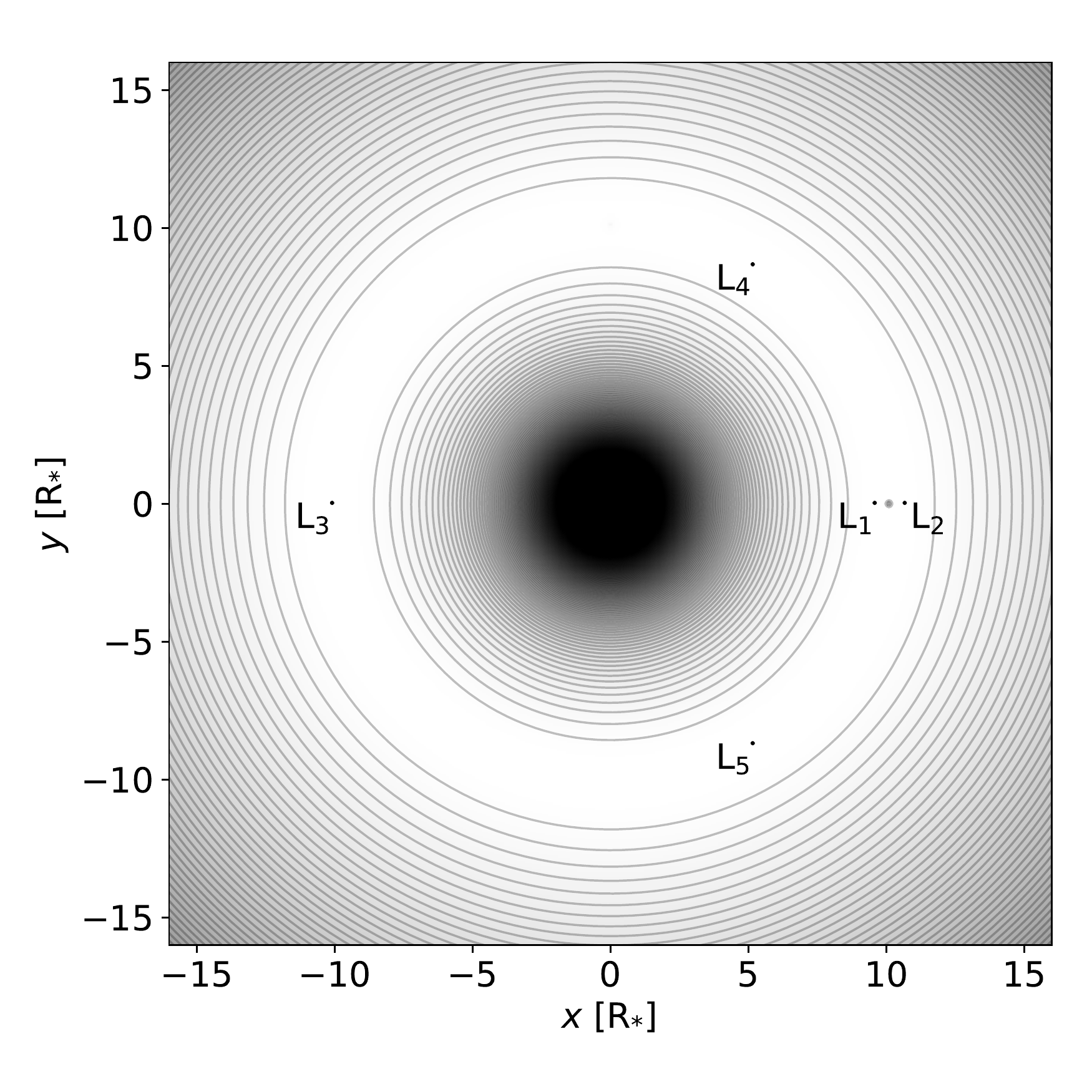}
	\includegraphics[width=0.49\textwidth]{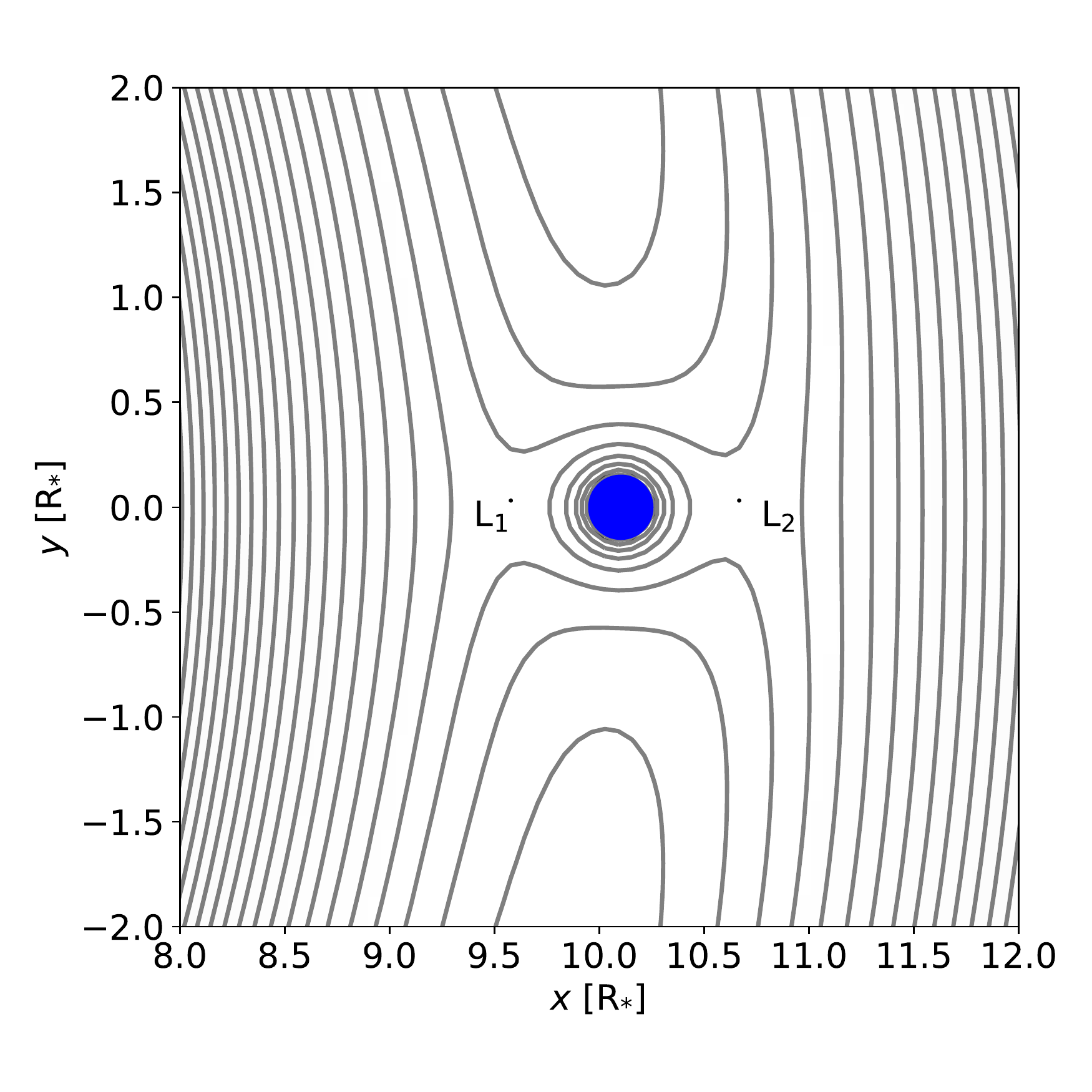}
	\caption[Roche equipotential for both the system and hot Jupiter vicinity.]{Left: Roche equipotential for the inner portion of the system. At the centre is the star and to the right the planet, with Lagrange points 1 - 5 indicated. Right: close up view of the Roche equipotential for the circumplanetary environment with the blue circle indicating the radius of the HJ. Lagrange points L$_{1}$ and L$_{2}$ are shown to the left and right and their distances from the planetary surface are $2.48 \ R_{\circ}$ and $2.75 \ R_{\circ}$ respectively.}
	\label{ACCRETIONfig:Roche}
\end{figure*}

To gain an insight into the gravitational topology of the HJ system, the Roche equipotential is plotted in Fig. \ref{ACCRETIONfig:Roche}. If we consider only gravitational effects and ignore any fluid action from the stellar wind, for example wind ram pressure or mixing due to instabilities, we can build a picture of the planetary wind dynamics which we can contrast to the more sophisticated MHD simulation presented in later sections.
	
The $L_{1}$ point is only $2.48 \ R_{\circ}$ from the planetary surface and if we assume material leaving the planetary surface is traveling with a speed at least that of the planetary escape velocity, $v_{\mathrm{escp}, \circ}~=~34.85 \ \mathrm{km/s}$, material leaving the planetary surface will reach the $L_{1}$ point in at most $2.07 \ \mathrm{hours}$ or $0.023$ orbits. This tells us that mass lost from the day-side of the planet will enter under the dominant influence of the stellar gravitational field almost immediately. Material from the night-side will reach the $L_{2}$ point, situated at $2.75 \ R_{\circ}$ from the planets surface, in at most $2.30 \ \mathrm{hours}$ or $0.026$ orbits, again leaving the planets gravitational influence in a fraction of an orbit and moving out beyond the orbital radius. Material leaving the planet then falls into orbit around the star where it interacts with the stellar wind via ram pressure, mixing via fluid instabilities and magnetic forces.



\subsection{Global evolution}
\label{ACCRETIONsec:global}

First the global evolution of the models will be analysed to asses the extent to which the initial conditions have been dissipated and whether the simulation has reached quasi-steady state.

\subsubsection{Accretion rate}
\label{ACCRETIONsec:accretion_rate}

To determine whether this steady state has been reached and also to characterize the deposition of material onto stellar surface, we calculate the flux through a surface, $S$, of a sphere centred around the star. For simplicity, we assume a spherical coordinate system with the three cardinal directions $r, \theta, \phi$. The flux is given by the integral over $S$ of the velocity vector projected onto the direction normal to $S$ and the density. Mass-flux is given by
\begin{equation}
\mathnormal{f}_{\mathrm{mass}} = \int_{\mathrm{S}} \rho \left( \theta, \phi \right)_{R} \boldsymbol{v}_{\bot} \left( \theta, \phi \right)_{R} \mathrm{d} S
\label{ACCRETIONeq:mass_flux}
\end{equation}
The subscript $_{R}$ means the quantity is evaluated at the radius, $R$, of the sphere whose surface is $S$. $\rho$ is the mass per unit volume and $\boldsymbol{v}_{\bot}~=~\boldsymbol{v} \cdot \boldsymbol{n}$ is the component of the velocity in the direction normal to $S$.

\begin{figure*}
	\centering
	\includegraphics[width=0.49\textwidth,trim={8cm 0.75cm 8cm 1cm},clip]{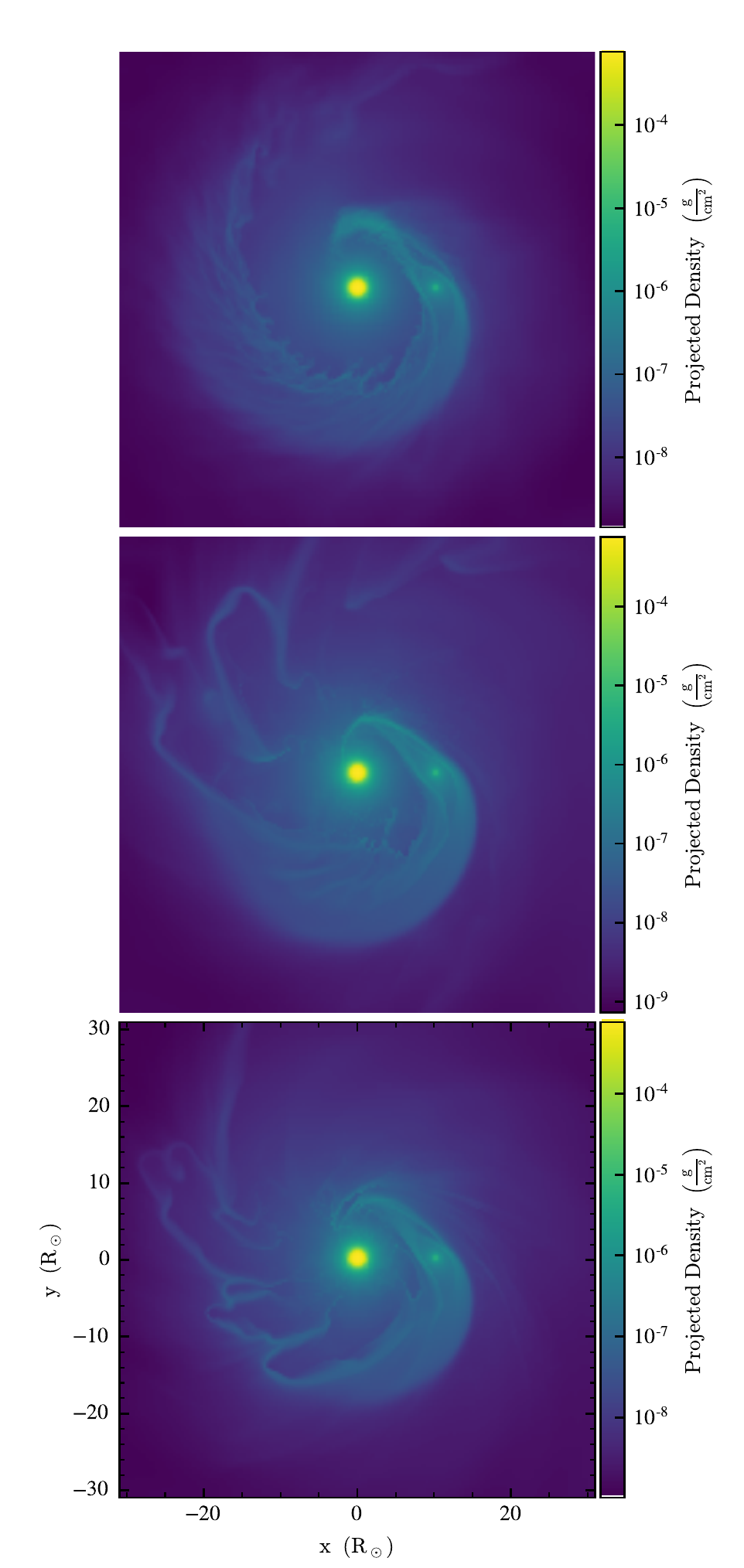}
	\includegraphics[width=0.49\textwidth,trim={8cm 0.75cm 8cm 1cm},clip]{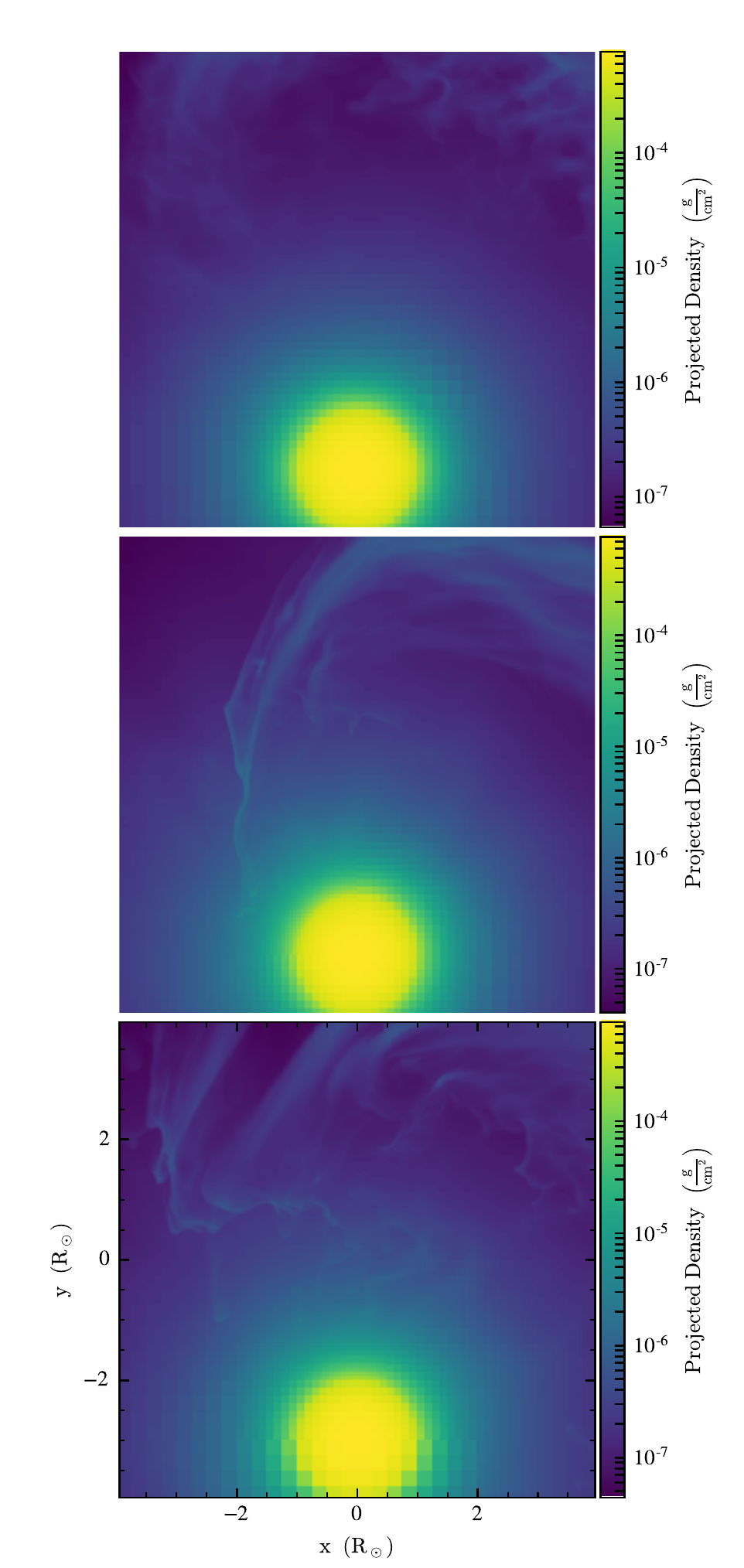}
	\caption[Large and small scale projection plots of the interacting hot Jupiter and stellar winds.]{Left column: projection plot of the large scale structure of the star-planet interacting wind. Right:  projection plot of the region directly ahead of the planets orbit including the stellar surface. Accreting material needs to pass through this region to reach the stellar surface, structures which develop here determine the morphology of the accretion stream. The three columns show, from top to bottom, data from models \textbf{S0P0}, \textbf{S2P1} and \textbf{S2P0} respectively.}
	\label{ACCRETIONfig:slice}
\end{figure*}

Fig. \ref{ACCRETIONfig:slice} shows a projection of the $z$ axis into the $xy$ plane of the simulation domain for all three models. Both the large scale structure of the star-planet interacting wind is shown on the left and on the right, a close up of the region directly ahead of the planets orbit which includes the stellar surface. This is an important region, as accreting material needs to pass through this area to reach the stellar surface and the morphology of developing structures here determines the rate and spatial location of accretion. The three columns show, from top to bottom, projections from models \textbf{S0P0}, \textbf{S2P1} and \textbf{S2P0} respectively.

Comparing the three models, only \textbf{S2P1} has developed a coherent accretion stream where material reaches the stellar surface, predominately through one conduit, only breaking up when reaching the inner magnetosphere. Model \textbf{S2P0} exhibits similar filamentary structure but has not developed into a single stream. Model \textbf{S0P0} has not developed any coherent streams through which material is transferred to the surface, the differing physical regime under which the simulation is conducted, HD rather than MHD, sets \textbf{S0P0} aside from the other two models. 

The planetary wind rapidly expands spherically to fill the planets Roche lobe and continues until it collides with the stellar wind. All three models develop a tear-drop like structure around the HJ, this structure is a result of the balance between outward ram pressure of the planetary wind and the large scale stellar wind. Beyond this, the planetary wind is forced to collimate into a stream ahead of the planet where it spirals into the inner stellar magnetosphere, and behind the planet where it expands in an open fan like structure until it reaches the outer boundary of the simulation.

For these three models, \textbf{S2P1}, \textbf{S2P0} and \textbf{S0P0}, only the first, \textbf{S2P1}, forms a coherent accretion stream. The material forming the steam itself can only be followed to within $1.5 \ R{_\ast}$ of the stellar surface, as the simulation in this region is held constant to ensure the wind is initialised properly (see Section \ref{ACCRETIONsec:BCs}). To overcome this, model \textbf{S2P1} was also simulated using a statically refined spherical polar grid, allowing material to be traced all the way down to the stellar surface. The next sections details the results of both the Cartesian and spherical polar version of model \textbf{S2P1}, starting with the mass lost by the two bodies and a consideration of the possible transfer of mass from one body to the other.

\subsection{Mass-loss}
\label{ACCRETIONsec:global_mass_flux}

\begin{figure}
	\centering
	\includegraphics[width=0.49\textwidth]{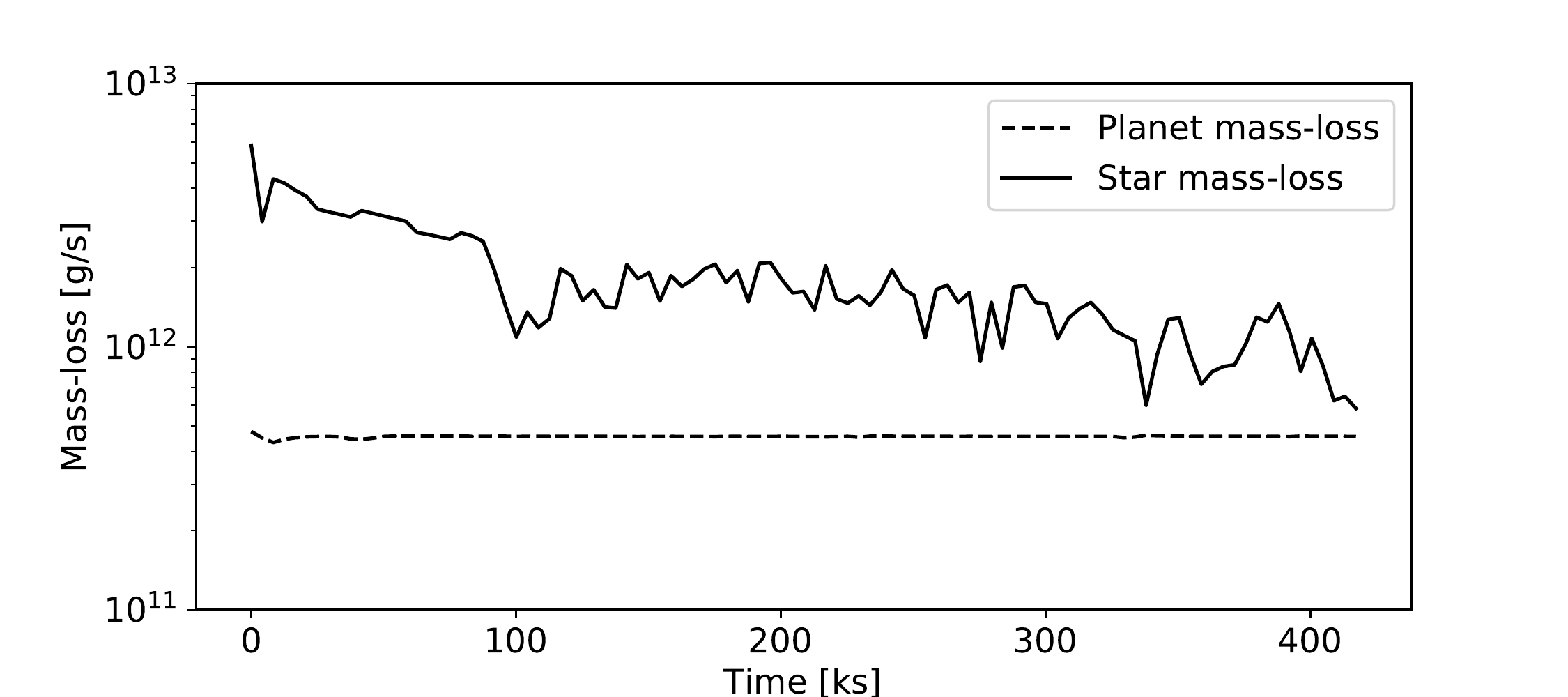}
	\caption[Stellar and hot Jupiter mass-loss rates for the accretion simulation.]{Mass-loss as a function of time for both the star and HJ in the Cartesian version of model \textbf{S2P1}. Mass-loss for the star, $\dot{M}_{\ast}$, undergoes a decrease from its initial value to $\sim10^{12} \ \mathrm{g/s}$. The curve exhibits fluctuations in its profile after $\sim100 \ \mathrm{ks}$, beyond this point material from the planet advects close enough to the star to influence the mass-loss calculation. In the HJs case, $\dot{M}_{\circ}$ is constant from the beginning of the simulation at a value of $4.6~\times~10^{11} \ \mathrm{g/s}$.}
	\label{ACCRETIONfig:flux_mass}
\end{figure}

Mass-loss from the star are determined by evaluating equation (\ref{ACCRETIONeq:mass_flux}) over a surface $S$ encompassing the star. We perform the analysis for the planetary mass-loss in the same manner and the results for both are plotted in Fig. \ref{ACCRETIONfig:flux_mass}.

The stellar mass-loss profile exhibits fluctuations after $\sim100 \ \mathrm{ks}$ which continue to the end of the simulation with  $\dot{M}_{\ast}~\sim10^{12} \ \mathrm{g/s}$. in the HJ case $\dot{M}_{\circ}$ is virtually static throughout the simulation at $\dot{M}_{\circ}~=~4.6~\times~10^{11} \ \mathrm{g/s}$. The explanation for the lack of fluctuating HJ mass-loss is that, within the immediate vicinity of the HJ, the planetary wind is entirely dominant and radially outward, with no material form the stellar wind approaching the surface $S$ where $\dot{M}_{\circ}$ is calculated.

Calculation of both $\dot{M}_{\ast}$ and $\dot{M}_{\circ}$ is conducted above the stellar or planetary surface to avoid boundary condition effects. Therefore, the fluctuation in the $\dot{M}_{\ast}$ profile does not necessarily correspond to planetary material accreting directly on to the stellar surface (see in Section \ref{ACCRETIONsec:surface} for the calculation of this quantity). Material can be swept back by the stellar wind before making contact and the stellar wind can be stalled by the incoming HJ material. As the magnitude of the fluctuations are not of the order of the HJ mass-loss rate, there is some other form of iteration accounting for this variability. 

\subsection{Inner magnetosphere}
\label{ACCRETIONsec:InnerMag}

To assess the exact position and extent of the accretion in model \textbf{S2P1}, a spherical polar grid is employed allowing the stellar surface to be a dynamic region in the simulation rather than a static boundary as is the case in the Cartesian version of the model. 

\begin{figure*}
	\centering
	\includegraphics[width=0.99\textwidth]{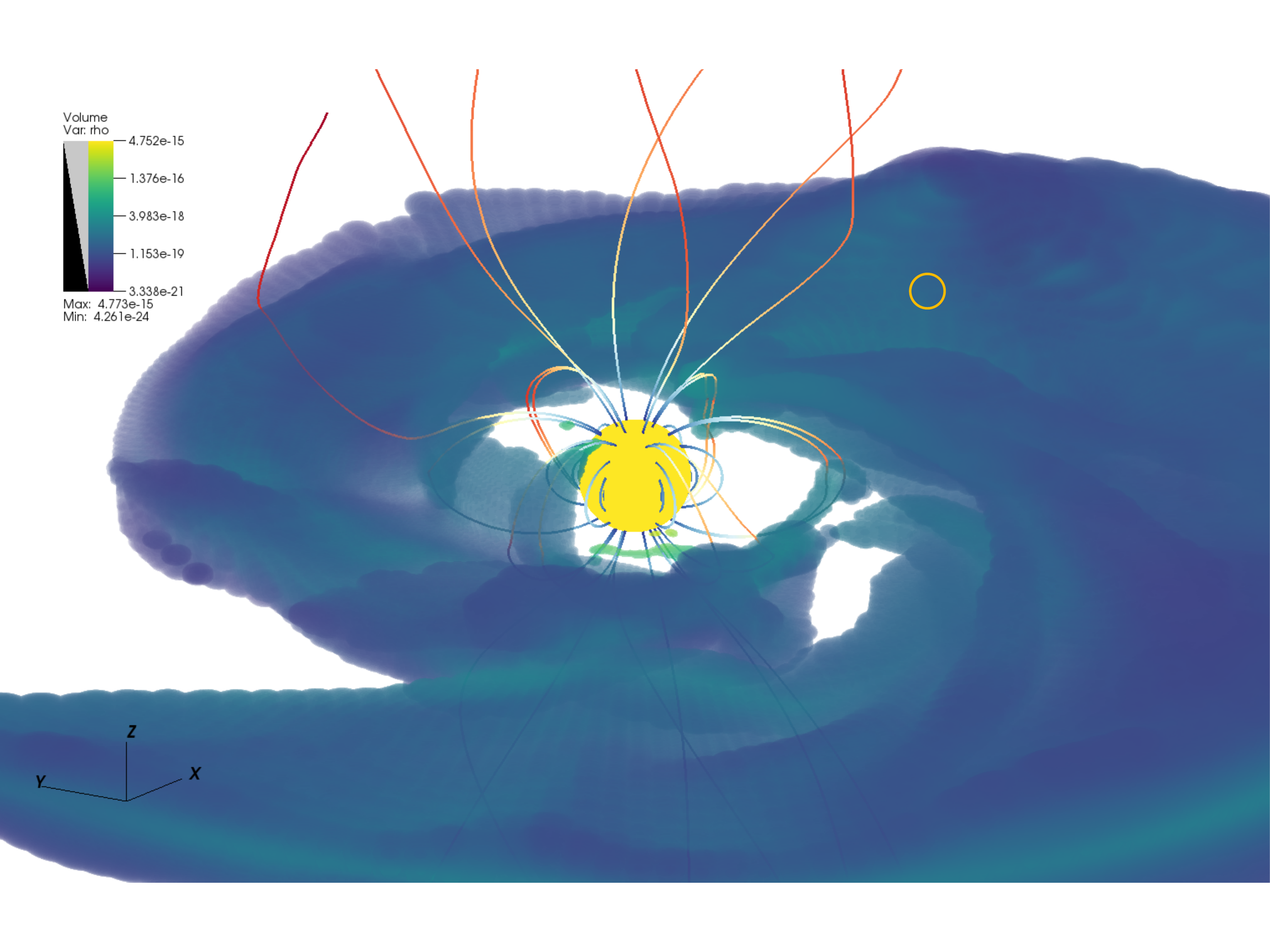}
	\caption[Volume rendering of the interaction between planetary mass-loss and the magnetic field of the host star.]{Volume rendering of the interaction between planetary material where $n_{\mathrm{pass}} > 1\%$ and the magnetic field of the host star. The planet is embedded in the cloud of material in the background and is indicated by the annotated orange circle, behind the star which is the yellow sphere at the centre of the image. Magnetic field lines originating in the polar regions of the stellar surface are drawn open and contorted by the wind to form a spiral structure, while field lines in the equatorial region remain closed. The field line colour scheme represents the strength of the magnetic field, from stronger field (blue) to weaker field (dark red).}
	\label{ACCRETIONfig:volume}
\end{figure*}

Fig. \ref{ACCRETIONfig:volume} illustrates the expansion of planetary atmospheric material into the inner magnetosphere of the star. By using the passive scalars to select only the planetary wind, where $n_{\mathrm{pass}} > 1\%$ of the planetary surface concentration, it can be seen that arcing material makes fall on the stellar surface ahead of the planet's orbit. The planet can not be seen in this image as it is embedded in the cloud of material in the background, behind the star which is the yellow sphere at the centre of the image.

\begin{figure*}
	\centering
	\includegraphics[width=0.79\textwidth]{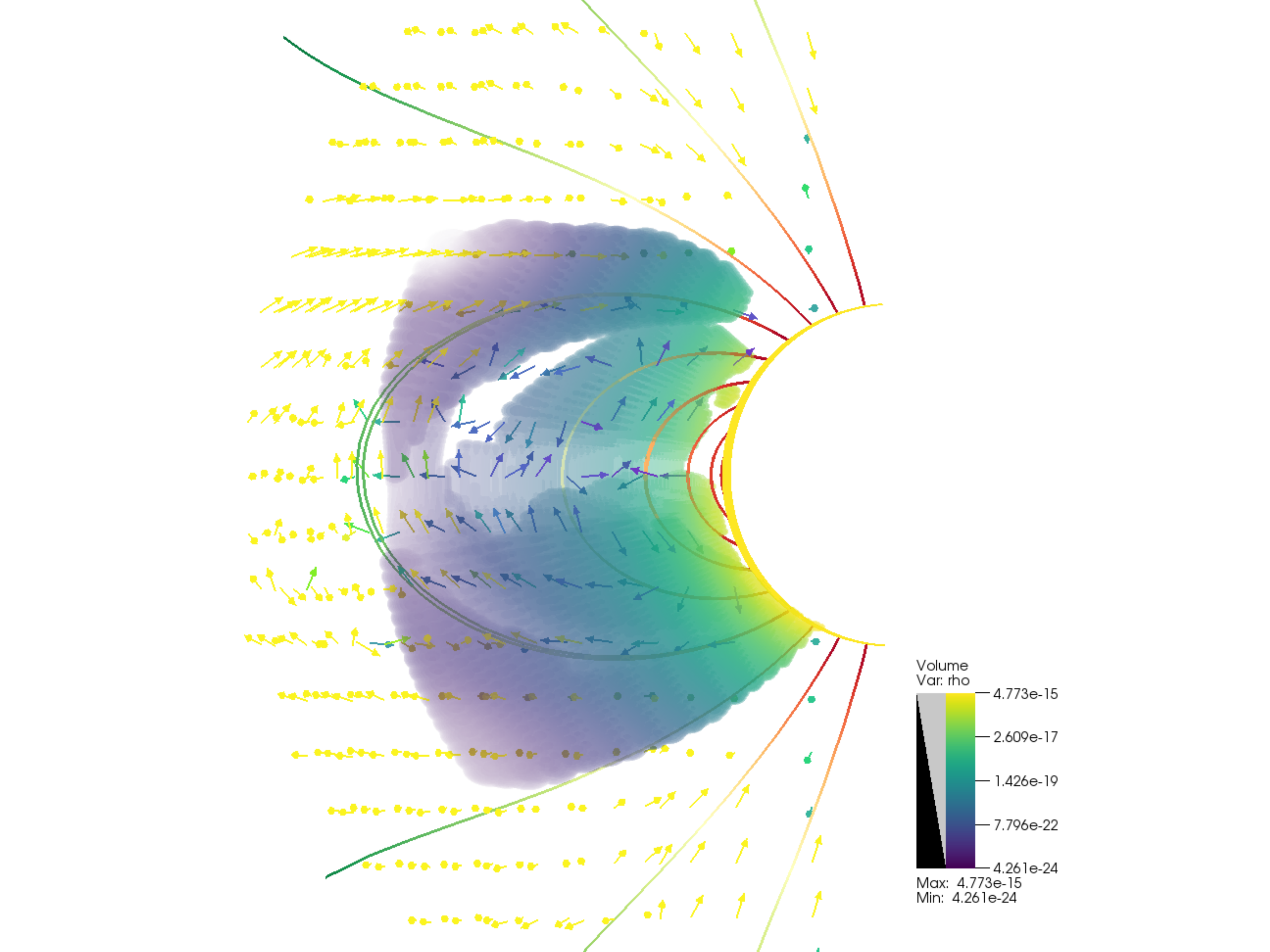}
	\caption[Volume rendering of planetary material in the magnetosphere of the star with field lines and velocity vectors.]{Volume rendering of the density of planetary material suspended in the magnetosphere of the star for a thin region of width 5$^{\circ}$ centered at 220$^{\circ}$ ahead of the sub-planetary point. Magnetic field lines are coloured to indicate the magnitude of magnetic tension. Field lines close to the stellar surface exhibit the strongest tension and therefore experience the smallest perturbation away from the initial dipole. The material structure and velocity vectors can clearly be seen to form arcs corresponding to the field lines, with arrows pointing along filed lines and not purely radial as would occur in the non-magnetized case. This indicates that material close to the stellar surface is dominated by magnetic pressure. What is shown is material where the $n_{\mathrm{pass}}$ concentration is above 1\%.}
	\label{ACCRETIONfig:tension}
\end{figure*}

Stellar magnetic field lines are either drawn open and contorted by the wind to form a spiral structure or remain closed and retain their dipolar structure. The division between these two regions is a function of the stellar wind ram-pressure and strength of the stellar dipole. At equilibrium, there exists a latitudinal angle specifying the transition between open and closed field lines and is given by 
\begin{equation}
\sin^2 \left( \Theta_{\mathrm{m}} \right) = \frac{R_{\ast}}{r_{\mathrm{m}}}.
\label{ACCRETIONeq:angle_latitude}
\end{equation}
Where $R_{\ast}$ is the stellar radius, $r_{\mathrm{m}}$ is the radial distance from the centre of the star to the apex of the longest closed field line and $\Theta_{\mathrm{m}}$ is the corresponding angle between the $z$-axis and the point where the longest field line makes contact with the surface \citep{Jardine2006}. Equation (\ref{ACCRETIONeq:angle_latitude}) assumes no deformation of the field lines due to either stellar rotation or gas motion, an assumption which is largely accurate in the innermost region ($< 1.5 R_{\ast}$ above the stellar surface) of the magnetosphere (see below). A physically motivated estimate for $r_{\mathrm{m}}$ is needed in order to use equation (\ref{ACCRETIONeq:angle_latitude}). \cite{Jardine2006} describe a model in which gas pressure is balanced with magnetic pressure to provide a parameterised estimate for $r_{\mathrm{m}}$ for a T Tauri stellar magnetosphere (although the model is general for stellar magnetospheres). In the context of the present study, we can estimate this radius via visual inspection of Fig. \ref{ACCRETIONfig:volume} and \ref{ACCRETIONfig:tension}. From this we will assume for simplicity $r_{\mathrm{m}} = 3 \ R_{\ast}$, yielding a latitudinal angle of $55^{\circ}$. This angle will be compared directly to the accretion latitude from the simulations.

Planetary material in Fig. \ref{ACCRETIONfig:volume} can be seen spiralling into the stellar magnetosphere from the planetary orbital radius, this material then interacts with the closed field lines where magnetic tension acts to support the material. It is then confined to follow these field lines down to the stellar surface. Therefore, the position where these closed field lines intersect the stellar surface determines the exact location where planetary material makes foot-fall and therefore the exact surface position at which accretion takes place (see Section \ref{ACCRETIONsec:surface}).

To further investigate the interplay between the accreting planetary material and the magnetic field of the star, we plot the planetary material and magnetic field lines in a region directly above the stellar surface. Fig. \ref{ACCRETIONfig:tension} shows this interplay. Magnetic field lines are coloured according to the strength of magnetic tension, given by $| \left( \mathbf{B} \cdot \nabla \right) \mathbf{B}/4\pi |$, which acts to restore the magnetic field lines to the lowest energy configuration. In the case of a pure dipole, the magnetic tension is balanced by magnetic pressure resulting in a force free field. For the star in our simulations, this dipole structure is largely preserved out to the Alfv\'{e}n surface. Beyond this, the field is drawn open by the expanding stellar wind.

Stellar rotation also acts to warp field lines. This acts in the $\phi$-direction and laterally stresses the field line, as can be seen in Fig. \ref{ACCRETIONfig:tension} and at a larger scale in Fig. \ref{ACCRETIONfig:volume}. This effect grows as the radial distance form the star increases, as field lines which protrude further into the wind are azimuthally contorted to a greater degree. Velocity arrows are annotated in Fig. \ref{ACCRETIONfig:tension} to illustrate the direction of material. In the equatorial region, the velocity is directed along the field lines.

The net effect of the stellar magnetic field is to guide the incoming planetary material towards the stellar poles rather than to accrete directly onto the equatorial region. This is consistent with the frozen in condition of ideal MHD. increasing the stellar dipole strength will lead to accretion at higher latitude and inversely, weaker dipole strength will lead to accretion at lower latitude, in agreement with equation (\ref{ACCRETIONeq:angle_latitude}). If non-ideal plasma physics are incorporated, such as resistivity or other dissipative mechanisms, then a departure from this behavior may be seen. 

Our results are consistent with simulations conducted by \cite{Long2007} who investigated the accretion of protoplanetary disc material onto the magnetospheres of dipole and non-dipole early type low mass stars. \cite{Long2007} find that the accretion occurs in predominately ring like patterns, parallel to the equator or individual spots close to the poles. The precise pattern of accretion which occurs in our study is described in the next section. 

\subsection{Stellar surface evolution}
\label{ACCRETIONsec:surface}

In the following sections, when referring to locations on the stellar surface, latitude and longitude are given as $\theta$ and $\phi$ and a distinction is made from the spherical grid coordinates used in Section \ref{ACCRETIONsec:Num}, by following the angle with compass directions. For example, the position $\theta~=~ 30^{\circ}$N and $\phi~=~40^{\circ}$E refers to the point on the stellar surface which is both $30^{\circ}$ above and $40^{\circ}$ ahead of the subplanetary point.

\begin{figure*}
	\centering
	\includegraphics[width=0.99\textwidth]{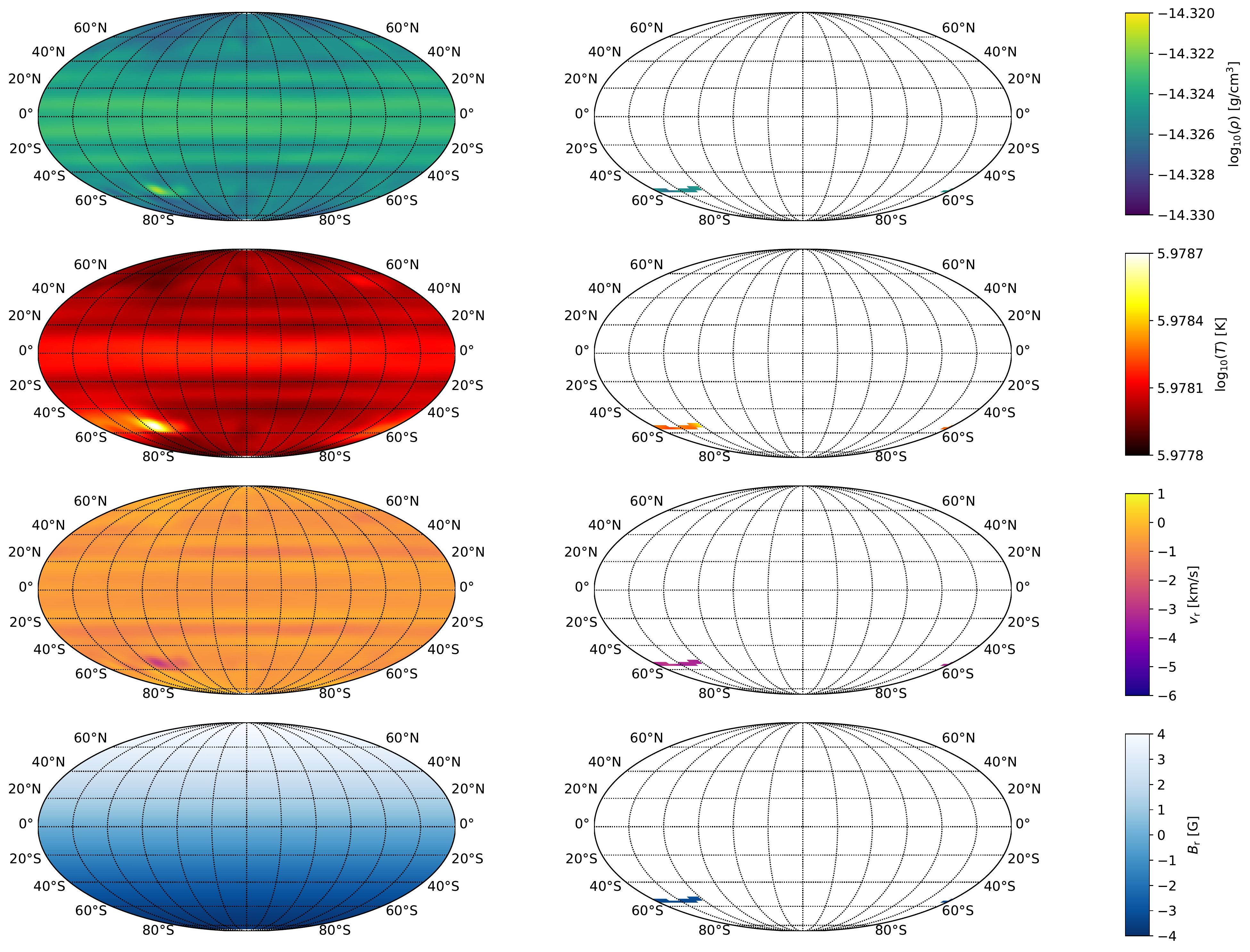}
	\caption[Surface plots of the star showing both the complete surface variability and the accreation spot.]{Left column: stellar surface values for $\rho$, $T$, $v_{\mathrm{r}}$ and $B_{\mathrm{r}}$. This column shows the total material, both from the stellar and planetary winds. Right column: planetary wind values only; the regions displaying data are where $n_{pass}$ are present. Lines of latitude and longitude mark equal distances on the stellar surface ($20^{\circ}$ in latitude and $30^{\circ}$ in longitude). Zero latitude and zero longitude corresponds to sub-planetary position, the centre of each plot. The extrema of longitude are the opposite side of the star. Not only do the $n_{pass}$ show what the magnitude of the accreting quantities, but also the spatial location at which the accretion takes place; $\phi~=~133^{\circ}$W (or $227^{\circ}$ ahead of the sub-planetary point) and $\theta~=~53^{\circ}$S.}
	\label{ACCRETIONfig:ortho}
\end{figure*}

\begin{figure*}
	\centering
	\includegraphics[width=0.99\textwidth,trim={0 0 0 0},clip]{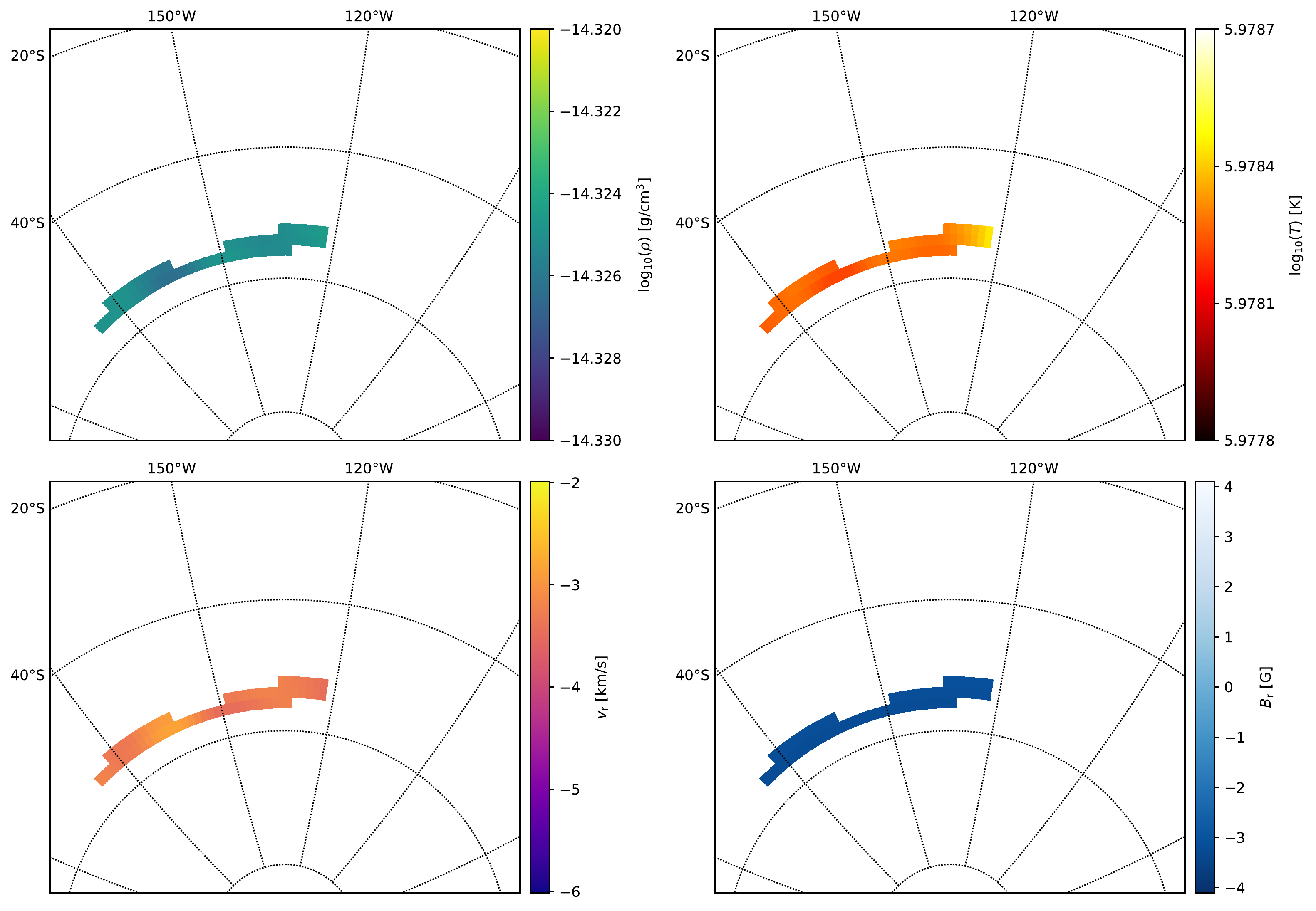}
	\caption[Close up of the accretion spot on the stellar surface.]{Close up of the accretion point located at $\phi~=~227^{\circ}$ ahead of the sub-planetary point and $\theta~=~-53^{\circ}$ below the orbital plane. Starting at top left and going clockwise, the plots are $\rho$, $T$, $v_{\mathrm{r}}$ and $B_{\mathrm{r}}$. The accretion point resembles a narrow strip and less like a spot. The cutoff for $n_{pass}$ is conservative and one should assume that the accretion region extends beyond this strip, however, we exclude regions where the $n_{pass}$ concentration is less than 1\% of the planetary surface concentration.}
	\label{ACCRETIONfig:ortho_close}
\end{figure*}

Fig. \ref{ACCRETIONfig:ortho} and \ref{ACCRETIONfig:ortho_close} depict the stellar surface at time $t~=~417.3 \ \mathrm{ks}$ (the final time of the simulation), in orthographic form for both the complete surface and the region in which planetary $n_{pass}$ are present. $\rho$, $T$, $v_{\mathrm{r}}$ and $B_{\mathrm{r}}$ show the impact of the accreting material, resulting in a departure from the ambient stellar surface conditions. The location of accretion is displayed in the right hand column. Here, the accretion point is located at approximately $\phi~=~133^{\circ}$W, or $\phi~=~227^{\circ}$ ahead of the sub-planetary point and $\theta~=~53^{\circ}$S. This shows that the accretion stream loops almost completely around the star before making contact. This stellar surface position differs from results found by \cite{Pillitteri2015} who conducted a limited simulation to provide a physical mechanism in support of observational evidence for SPMI. In their study, an accretion stream, which makes contact with the stellar surface at $\phi~\sim~70^{\circ}$ ahead of the sub-planetary point, is described in detail including a knee like structure where the stream is swept back in, counter to the orbital direction. This is due to the slower rotation of the star compared to the orbital rotation of the planet they simulate. However, as the star in our simulation rotates at the same frequency as that of the orbit, the inner accretion stream does not experience the drag which \cite{Pillitteri2015} describe and no knee feature develops, as can be see in the features of the right hand column of Fig. \ref{ACCRETIONfig:slice}. This indicates that the precise location of the accretion spot is strongly influenced by the rotation of the star itself. A more detailed comparison with this study is discussed in Section \ref{ACCRETIONsec:implications}.

The largest perturbation to the stellar surface quantities actually occurs ahead of the accretion point, outside the region marked by the passive scalers. This feature has the highest density and temperature and lowest radial velocity on the stellar surface. A possible explanation is that as the accretion stream approaches the stellar surface it pushes ahead of it a bow shock of stellar wind material and that the most significant perturbation to the surface is in fact of stellar origin via interaction with HJ material.

As previously stated, the highest concentration of accreting material is found $\phi~=~227^{\circ}$ ahead of the sub-planetary point and $\theta~=~53^{\circ}$ below the orbital plane. This value is consistent with the calculated value of $\Theta_{\mathrm{m}}~=~55^{\circ}$N/S (either north or south) from Section \ref{ACCRETIONsec:InnerMag}, equation (\ref{ACCRETIONeq:angle_latitude}) which assumed $r_{\mathrm{m}}~=~3 \ R_{\ast}$.

\begin{figure}
	\centering
	\includegraphics[width=0.49\textwidth,trim={0 6cm 0 6cm},clip]{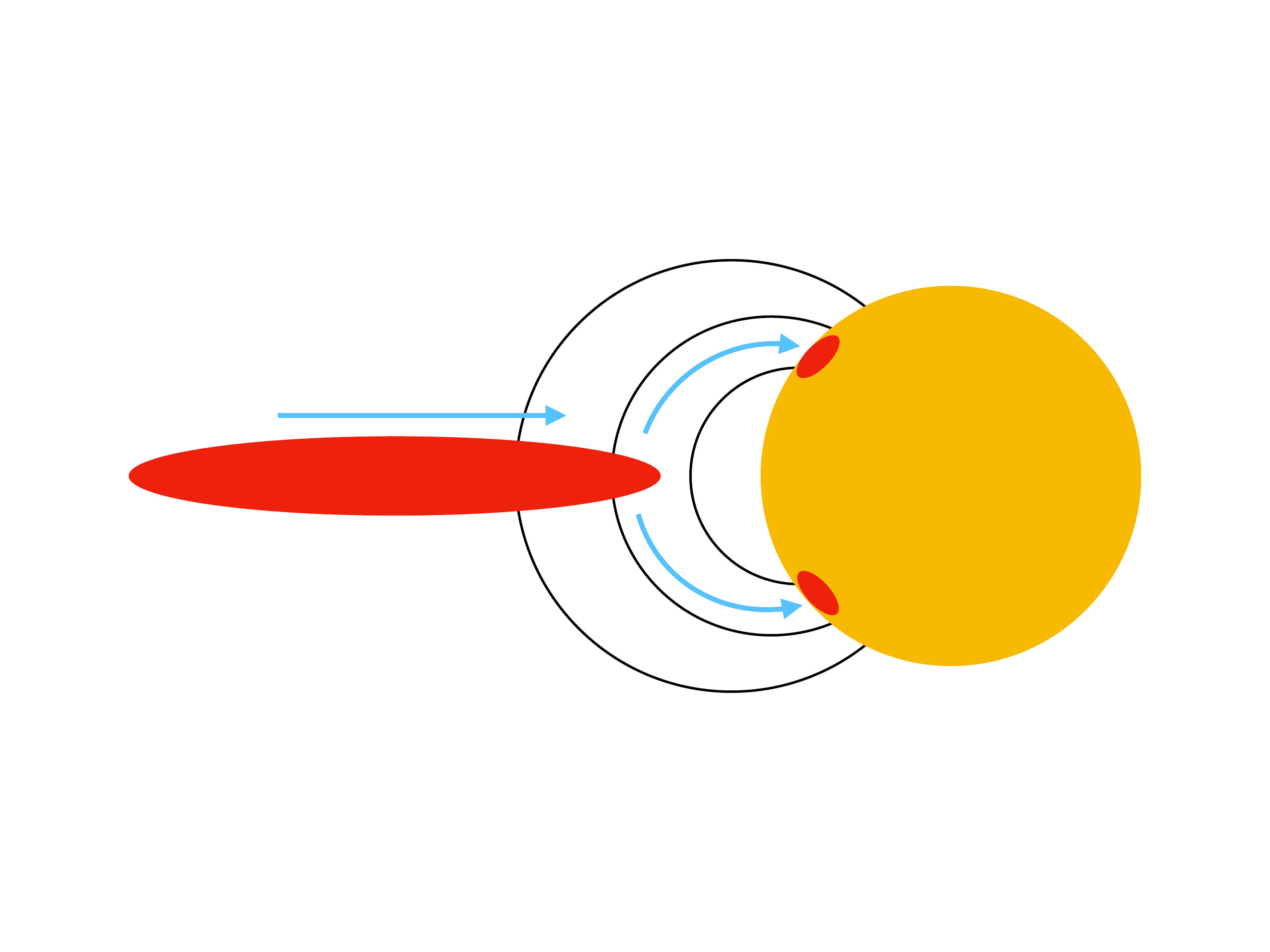}
	\caption[Cartoon illustrating the motion and position of incoming accreting material.]{Cartoon illustrating the motion and position of incoming accreting material. The thin accretion stream makes its way into the outer magnetosphere  remaining coherent until interacting with the closed field lines close to the stellar surface, here the magnetic tension consts the incoming material to follow magnetic field lines down to the stellar surface. The precise location of which will depend on the local plasma-$\beta$ value.}
	\label{ACCRETIONfig:cartoon}
\end{figure}

To further illustrate the trajectory of the accreting planetary material, Fig. \ref{ACCRETIONfig:cartoon} shows a cartoon diagram of the motion as it enters the stellar magnetosphere and interacts with the field lines for model \textbf{S2P1}. The thin accretion stream remains coherent as it approaches the stellar surface, at a given point the gas ram pressure of the accretion stream is balanced by the magnetic pressure of the magnetosphere and stellar wind ram pressure. At this point material is confined to follow the stellar magnetic field lines down to the surface, resulting in the accretion spot seen in Fig. \ref{ACCRETIONfig:ortho} and \ref{ACCRETIONfig:ortho_close} at the latitude given by equation (\ref{ACCRETIONeq:angle_latitude}). 

Both Fig. \ref{ACCRETIONfig:cartoon} and equation (\ref{ACCRETIONeq:angle_latitude}) predict that accretion should take place on both the northern and southern hemispheres simultaneously, they are symmetric about the equator. As we only see accretion on the southern hemisphere in our simulation we can conclude that the symmetry is broken by dynamic properties of the MHD flow.

The coordinates of the accretion spot stated above are instantaneous and capture no information about the intermittent nature of the accretion or how its spatial location changes. To determine this, both the flux of accreted material and the accretion coordinates are plotted as a function of time. This allows us to make quantitative statements about the impact of the accretion stream on the stellar surface and possible observable signatures. 

To achieve this, we take the ambient surface condition from the early stages of the simulation, when the stellar surface has relaxed from a Parker wind to a magnetospheric configuration but no planetary material has yet interacted with the surface. This average, ambient stellar surface is then used to normalise the density and temperature in the accretion spot resulting in an over or under density and temperature of the spot, giving the net effect of the planetary accretion stream. The normalised values can therefore be thought of as the percentage increase (or decrease) in surface fluid quantities due to the action of accreting planetary material. 

\begin{figure}
	\centering
	\includegraphics[width=0.49\textwidth,trim={0 6cm 0 7cm},clip]{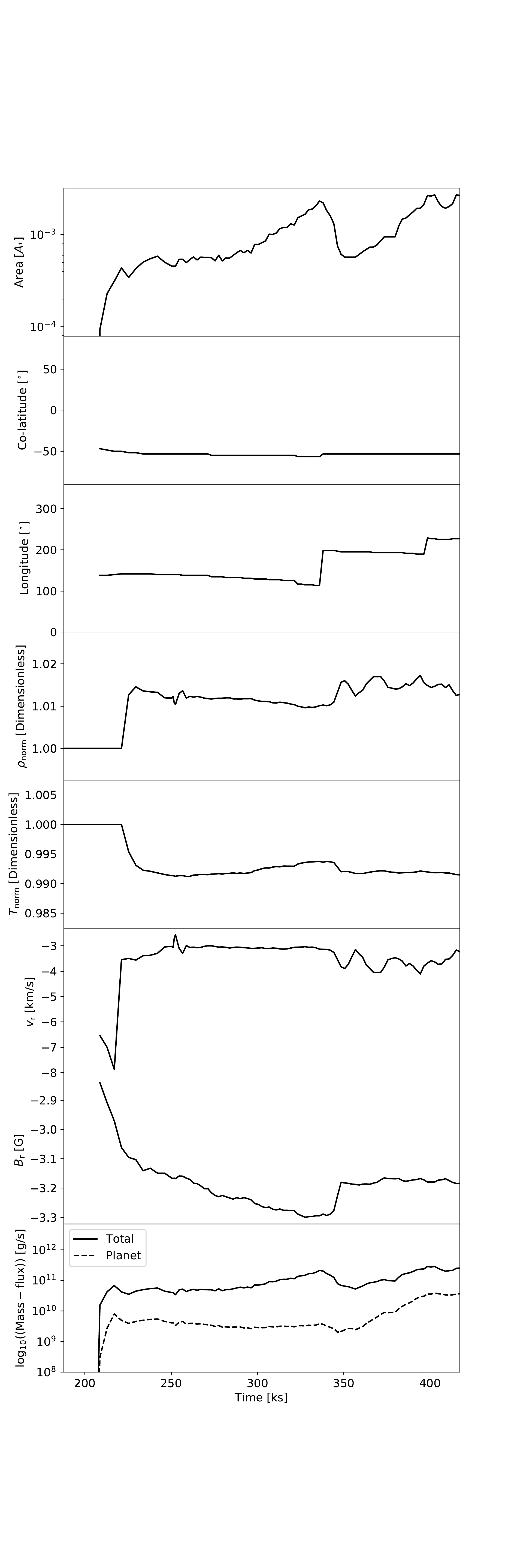}
	\caption[Accretion spot size, location on the stellar surface, fluid quantities and mass-flux through spot.]{From top to bottom, accretion spot size, location, fluid quantities $\rho$, $T$, $v_{\mathrm{r}}$, $B_{\mathrm{r}}$ and Mass-flux through spot. Accretion begins at $207 \ \mathrm{ks}$. $\rho$ and $T$ are normalised to the base, ambient values. The average quantities over the final $50 \ \mathrm{ks}$ of the simulation are: $\mathrm{Spot \ Area}~=~1.78~\times~10^{-3} \ A_{\ast}$, $\rho_{\mathrm{norm}}~=~1.02$, $T_{\mathrm{norm}}~=~0.99$, $v_{\mathrm{r}}~=~-~3.66 \ \mathrm{km/s}$, $B_{\mathrm{r}}~=~-~3.17 \ \mathrm{G}$. Finally, the mass-flux through the accretion spot is subdivided into the total mass-flux, $1.89~\times~10^{11} \ \mathrm{g/s}$, and the mass-flux of planetary material, $2.38~\times~10^{10} \ \mathrm{g/s}$.}
	\label{ACCRETIONfig:Average_spot_quantities}
\end{figure}

Beginning at $207 \ \mathrm{ks}$, when accretion commences, the evolution of the spot latitude, longitude, size and the normalised quantities, along with the average radial velocity and magnetic field in the spot are plotted in Fig. \ref{ACCRETIONfig:Average_spot_quantities}.

From Fig. \ref{ACCRETIONfig:Average_spot_quantities} we can make some attempt to quantify the stability of the accretion stream. We can see that while the latitude of accretion is stable at $\sim 50^{\circ}$S, the longitude processes round the star from $\sim 138^{\circ}$ to $\sim 227^{\circ}$ ahead of the sub-planetary point. The spot size undergoes a periodic variation between $5~\times~10^{-4} \ A_{\ast}$ and $3~\times~10^{-3} \ A_{\ast}$, where $A_{\ast}$ is the stellar surface area, indicating a pulsing in the accretion rate and a periodicity in the stream stability of $67 \ \mathrm{ks}$ (taking the separation in time of the two maxima). This variation however only last for two maxima and one minima before the simulation reaches its maximum time, as such it is unclear whether this will continue. 

The average quantities over the final $50 \ \mathrm{ks}$ are: $\mathrm{spot \ area}~=~1.78~\times~10^{-3} \ A_{\ast}$, $\rho_{\mathrm{norm}}~=~1.02$, $T_{\mathrm{norm}}~=~0.99$, $v_{\mathrm{r}}~=~-~3.66 \ \mathrm{km/s}$, $B_{\mathrm{r}}~=~-~3.17 \ \mathrm{G}$ and mass-flux$~=~1.89~\times~10^{11} \ \mathrm{g/s}$. $T_{\mathrm{norm}}$ and $\rho_{\mathrm{norm}}$ are the spot temperature and density normalised by the stellar surface values without accretion; allowing us to determine the extent to which the stellar surface is perturbed by accretion. 

The spot temperature being $99\%$ of the stellar wind temperature is consistent with the threshold passive scaler concentration of $1\%$ used to isolate the accretion spot. As the simulation is quasi-isothermal, this indicates there is a degree of numerical mixing between the stellar and planetary winds. This together with the very small velocity values (in comparison to the stellar escape velocity $\sim 100 \ \mathrm{km/s}$) implies that the accreting HJ material slowly sinks rather than free-falls to the stellar surface.

\subsubsection{Mass accretion}

Of the above quantities, the spot size exhibits the largest variation over the course of the simulation, the average density remains approximately constant at $1\%$ above ambient density, implying that the accretion stream feeds material to the stellar surface at a constant density despite the variability in the size of the accretion spot. Together they result in the time-varying Mass-flux seen in the bottom plot of Fig. \ref{ACCRETIONfig:Average_spot_quantities} with the same period as the accretion spot size, though the amplitude is not as pronounced. 

The total mass-flux through the spot, averaged over the last 50 ks, is $1.89~\times~10^{11} \ \mathrm{g/s}$. If we compare this value to the mass-loss rates of both the star, $\dot{M}_{\ast}~=~9.49 \times 10^{11} \ \mathrm{g/s}$, and the HJ, $\dot{M}_{\circ}~=~4.60 \times 10^{11} \ \mathrm{g/s}$, (provided by Fig. \ref{ACCRETIONfig:flux_mass}) we find that the accretion spot mass-flux is 20\% and 41\% of these quantities respectively.

Not all of this material is purely from the planet however, as a high degree of mixing has occurred between the HJ and stellar winds. By using the passive scalars to weight the density field, we can select out the accretion spot mass-flux of purely planetary material. This mass-flux is $2.378~\times~10^{10} \ \mathrm{g/s}$. Again, if we compare this to the stellar and planetary mass-loss rates from Fig. \ref{ACCRETIONfig:flux_mass}, we find that it is 2.5\% and 5.2\% respectively. The two accretion spot mass-fluxes are displayed in the bottom plot of Fig. \ref{ACCRETIONfig:Average_spot_quantities}. 

These results tell us that, of the material lost from the planets atmosphere, $5.2\%$ is accreted onto the stellar surface. However, this material only comprises just over 1/10$^{\mathrm{th}}$ of the accretion spot mass-flux. The rest is stellar wind material, which has interacted with the planetary outflow and been drawn back down to the stellar surface.

The above ratio of planetary to stellar wind material in the accretion spot indicates that any enhanced stellar surface activity induced by the HJ may in fact be indirect and that stellar wind material is primarily responsible (As has been discussed at the beginning of Section \ref{ACCRETIONsec:surface}), having interacted and mixed with planetary material. This conclusion is given further credence if we consider the fact that the temperature and density within the accretion spot is approximately $\pm~1\%$ of the ambient conditions. This indicates that the stellar wind has stalled on the surface, in the presence of the incoming planetary material. The buildup of stellar wind material would naturally lead to an accretion spot dominated by stellar wind material, which is indeed what we find.

\subsubsection{Radial mass distribution}

\begin{figure}
	\centering
	\includegraphics[width=0.49\textwidth,trim={0 0 0 0},clip]{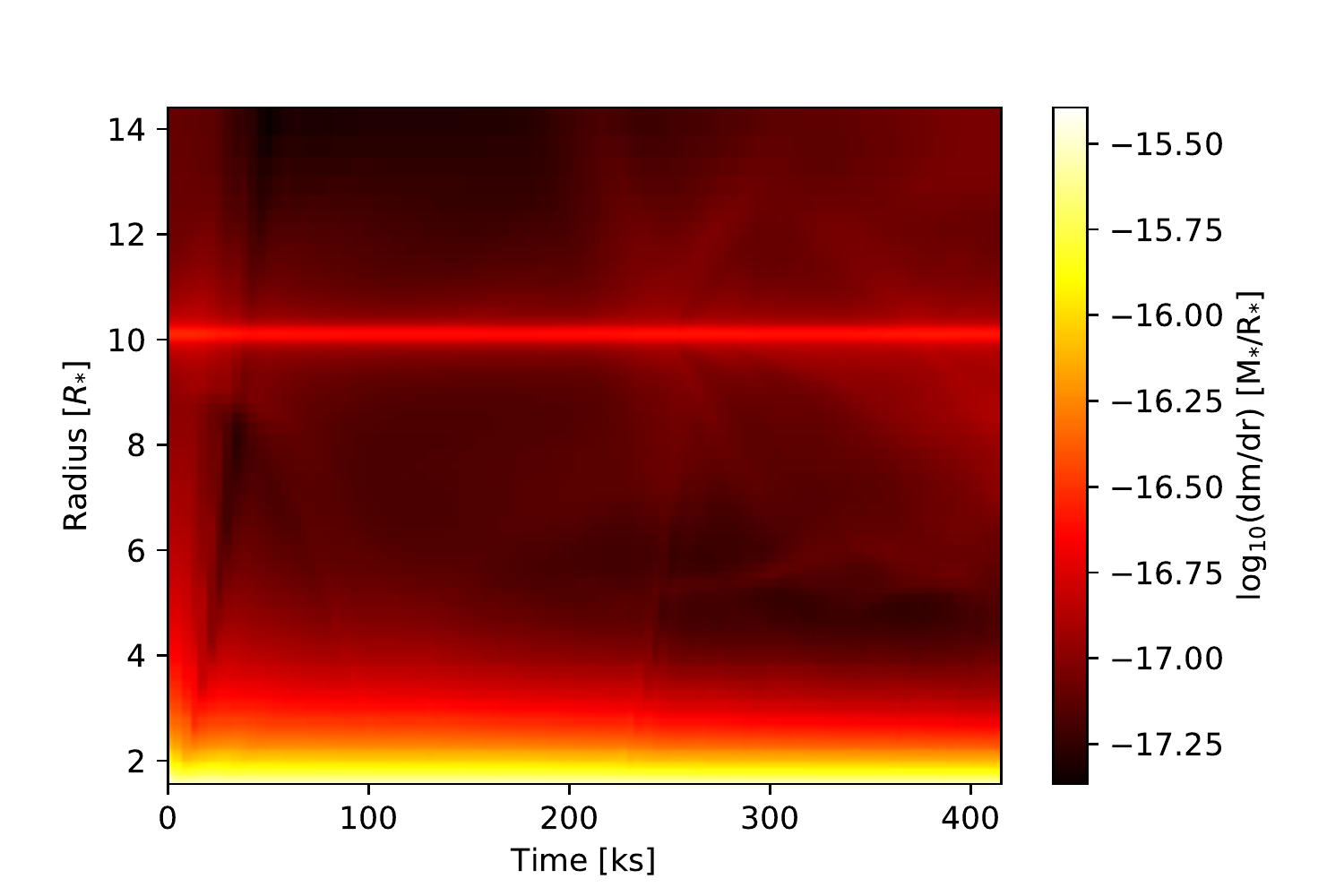}
	\caption[Time-dependency of the radial mass distribution for accreting planetary material]{Time-dependency of the radial mass distribution, $\mathrm{d} m \left( r, t \right) / \mathrm{d} r$. Two features at $1 \ R_{\ast}$ and $10 \ R_{\ast}$ are the stellar surface and planet respectively. The majority of the material in the system is stationary with respect to time. Faint features are however apparent and represent motion towards or away from the stellar surface.}
	\label{ACCRETIONfig:dmdr}
\end{figure}

Finally, to gain a global perspective of the motion of the accretion material described above, we use the approach of \cite{ud-Doula2008} and \cite{ud-Doula2013} to calculate the time-dependent radial distribution of mass via the following equation:
\begin{equation}
\frac{\mathrm{d} m \left( r, t \right)}{\mathrm{d} r} = r^{2} \int^{2 \pi}_{0} \int^{\pi}_{0} \rho \left( r, \theta, \phi, t \right) \sin(\theta) \mathrm{d} \theta \mathrm{d} \phi.
\label{ACCRETIONeq:dmdr}
\end{equation}
Here $r$, $\theta$ and $\phi$ are the three spherical polar coordinates and $\rho \left( r, \theta, \phi, t \right)$ is the density distribution. By integrating $\rho \left( r, \theta, \phi, t \right)$ first over $\theta$ between the limits $\frac{1}{4}\pi$ and $\frac{3}{4}\pi$ and then over $\phi$ in the full azimuthal range $0 \rightarrow 2 \pi$, we are left with $\mathrm{d} m \left( r, t \right) / \mathrm{d} r$ which is the radial distribution of mass for a given time $t$. by calculating equation (\ref{ACCRETIONeq:dmdr}) for each simulation snapshot, we build a picture of the global radial motion of mass. 
$\mathrm{d} m \left( r, t \right) / \mathrm{d} r$ is plotted in Fig. \ref{ACCRETIONfig:dmdr} in units of $\left[  M_{\ast}/R_{\ast}  \right]$. Any features whose radial distance changes with time are an indication of net flows either away or toward the stellar surface. Two features are immediately apparent at $1 \ R_{\ast}$ and $10 \ R_{\ast}$ and are the stellar surface and planet respectively. As most features remain at constant radius and there is no distinctive change at $207 \ \mathrm{ks}$, when accretion commences we can conclude that the majority of material is not involved in the accretion process. 

The next section will evaluate these results in the context of existing literature and observations.

\subsection{Implications for observable signatures}
\label{ACCRETIONsec:implications}

\cite{Cranmer2007} conducted theoretical work modelling SPMI between a HJ and a solar type star. Their work predicted that the precise nature of the accretion spot will vary from cycle to cycle if the star both rotates at a different frequency to the orbit and if the stars magnetic geometry is non-cylindrically symmetric, for example an oblique dipole or quadrupole, result in the predicted light curve not repeating exactly.

\cite{Shkolnik2008} report that SPI is intermittent and mimics cyclic variability of the host star and characterise this as the on/off nature of SPI. A striking result from the observations of \cite{Shkolnik2008} is that the leading longitude of activity seen on the stellar surface, attributed to the action of the HJ, is at a longitude of $\phi~\sim~70^{\circ}$ ahead of the sub-planetary point for HD 189733, HD 179949 and $\tau$ Boo, three out of their seven targets. For the latter, this phase is preserved over 5 yr implying steady SPI over this time-period. However, without knowing the mass loss rates of the targets, it is difficult to determine whether the form of SPI is due to direct magnetic connection between star and planet or due to an accretion stream.

Based on our results for the accretion spot location and the evolution of the spot size, the on/off nature of the planetary accretion described by \cite{Shkolnik2008} maybe attributed to the accretion stream either pulsing in size or precessing round the star to be out of phase with the planet's orbit or being obscured to the observer.

All the planets studied by \cite{Shkolnik2008} are orbiting within the Alfv\'{e}n surface of their host stars. The planets in our models are all orbiting outside the  Alfv\'{e}n surface and therefore in the super Alfv\'{e}nic region of the stellar wind. As such we classify our simulated systems as interacting via SPWI rather than SPMI. This is an important distinction as magnetic perturbations induced at the planetary orbit are unable to travel back to the stellar surface. This means that there is no heating of the stellar surface due to directly connected field lines between the star and the planet.

The HJ hosting system HD 189733 has been the focus of a number of studies in the context of SPMI \citep{Fares2010, Majeau2012, Pillitteri2015} observational and via MHD simulations. The system serves as an example of the complexity involved in studying SPMI systems. \cite{Fares2010} investigate the system using spectropolarimetry and reconstruct the magnetic map of the stellar surface with the result that it has a predominately toroidal surface magnetic field with a strength up to $40 \ \mathrm{G}$ (a departure from the simple dipole used in the present study). They conclude that stellar activity is mainly modulated by stellar rotation and find no evident of SPMI. 

For the HJ, HD 189733 b, \cite{Majeau2012} reconstruct a secondary eclipse map of the surface and allow them to deduce that the planet has both small obliquity and atmospheric winds which circulate in the atmosphere, described as \textit{super-rotating} winds.

More recently \cite{Pillitteri2015} studied this system with the Cosmic Origins Spectrograph on board Hubble Space Telescope (HST) and supported their observations with MHD simulation based on the work by \cite{Matsakos2015} and hence are similar in nature to those conducted here. Their observations show a high degree of variability in emission lines in Si, C, N, and O. They deduce that this enhanced activity is directly a consequence of the planetary material accreting onto the stellar surface. The point of accretion is inferred to be between longitudes $70^{\circ} - 90^{\circ}$ ahead of the sub-planetary point. Their simulations are however hampered by the numerical prescription which restricts the dynamic evolution of the simulation to be no closer than $1.5 \ R_{\ast}$ to the stellar surface and therefore the precise location of the accretion spot on the surface is unknown, a limitation not exhibited by the work presented here in the spherical polar version of model \textbf{S2P1}. The location of our accretion spot, $\phi~=~227^{\circ}$, does not agree with their reported position. The difference can be attributed to the fact that our planetary and stellar parameters are not the same as HD 189733 b and while a similar accretion longitude is reported for HD 179949 and $\tau$ Boo, there remains several factors which could lead to the difference. Firstly the rotational rate of HD 189733 is slower than the orbital rotation leading to the a swept back accretion stream as it interacts with the slower rotating stellar wind in the inner magnetosphere, a property our simulations lack due to the fixed rotational rate. Secondly, the shorter orbital period of the HD 189733 b makes the distance travelled by the accretion stream shorter and the stream would also have less angular momentum. This reduced arc distance would result in a trajectory for the accretion stream that would bring it close to the inner stellar magnetosphere just ahead of the sub-planetary point, as seen in their observations.

The SPMI occurring in HD 189733 is attributed to two different physical processes by \cite{Shkolnik2008} and \cite{Pillitteri2015}. For the former, SPMI is via direct magnetic field line connection between the stellar and HJ magnetospheres. Such interaction have been modelled by \cite{Strugarek2015a, Strugarek2016}. As has already been discussed, \cite{Pillitteri2015} attribute the interaction to an accretion stream from the HJ to the star, the same form of interaction as our models, SPWI.

The accretion spot quantities presented in Fig. \ref{ACCRETIONfig:Average_spot_quantities} provide the necessary detail to calculate the resulting observable signatures of the SPWI. The enhanced density and reduced temperature will lead to absorption and emission features that differ from those of the ambient  stellar surface. The precise enhancements due to the accretion stream we simulate will also increase the surface abundances of metals. The precise values will however be a function of the composition of both the stellar and HJ winds. However, the magnitude of the surface perturbation to both the density and temperature seen in the simulation presented here make detection unlikely. This is especially problematic for observations of spectral lines as the accretion spot temperature will result in ionisation of the heavier elements present in the HJ wind.

As our simulations only account for stellar surface increases due to accreting material, enhanced fine grained magnetic behaviour can not be captured and only large scale perturbations to the stellar dipole are present and, as we have discussed in Section \ref{ACCRETIONsec:InnerMag}, negligible. As such heating and cooling effects related to the action of the magnetic field are not captured and could account for the decrease in surface temperature in the accretion spot.  

Calculating specific observable signatures is however beyond the scope of this study, as our initial intention was to determine the nature of the accretion stream, specifically the stellar surface location and to demonstrate that a stable bridge of material between the two bodies can be formed in numerical simulations.

\section{Conclusions}

We have simulated a suite of mixed geometry, high resolution 3D MHD simulations which characterise the behaviour of interacting stellar and planetary wind material in the context of SPWI for a representative HJ hosting system.

Our results show that the pressence of a planetary magnetic field plays a central role in forming accretion streams between the star and HJ and that the nature of the accretion is variable both in location and in rate. The leading longitude of accretion in our simulation, $\phi~\sim~227^{\circ}$ ahead of the sub-planetary point, differs from reports in the literature which suggest a stable $\phi~\sim~70^{\circ}$ for a number of systems. This difference is attributed to specifics of the HJ system parameters. The accretion spot itself has been found to vary between occupying 0.05\% to 0.3\% of the stellar surface, a variation with a period of $67 \ \mathrm{ks}$. The mass-flux through the spot is $1.89~\times~10^{11} \ \mathrm{g/s}$ of which $12.6\%$ ($2.38~\times~10^{10} \ \mathrm{g/s}$) is planetary material, which in turn represents $5.2\%$ of the total material lost by the planet. Within the spot there is a 1\% decrease in temperature and 2\% increase in surface density. We predict this perturbation will lead to negligible observational signatures in spectral lines, as the temperature profile will ionise heavier elements. This is not in agreement with the observations of \cite{Pillitteri2015}. We have illustrated that magnetic fields cannot be ignored as the relatively modest field strengths used in our three models have lead to a dramatic difference in wind structure and determine the establishment of the accretion stream itself.

This difference in behaviour indicates that the establishment of an accretion stream is highly dependent upon the magnetic field of both the HJ and the host. While in all cases SPWI occurs and there are unique fluid features in the inter planetary medium due to SPWI, the accretion stream itself appears to require an aligned dipole-dipole combination. The authors acknowledge that the parameter space investigated in the present study is limited and further modelling is required to draw more precise conclusions.

The authors intend to further expand the theoretical basis for SPWI and the role stellar and planetary magnetic fields play in shaping its behaviour through the study of two specific HJ systems which directly compares SPMI with SPWI. This can be done by selecting a HJ whose orbit lies within its host's Alfv\'{e}n surface and one which lies outside.

There are now multiple HJ hosting systems exhibiting SPMI and little numerical modelling of specific systems. A concerted effort to systematically model these systems and deduce the stellar surface signatures of SPMI would allow for the classification and parameterisation of many HJ properties which remain ill constrained such as the nature of their magnetic fields.

\section{Acknowledgements}

The authors thank the reviewer for their helpful comments and suggestions; which improved the quality and content of the publication.

The authors acknowledge support from the Science and Facilities Research Council (STFC). 

Computations were performed using the University of Birmingham's BlueBEAR HPS service, which was purchased through HEFCE SRIF-3 funds. See http://www.bear.bham.ac.uk.

\bibliographystyle{mnras}
\bibliography{/Users/sdaley/Work/Reading/References/ExoplanetRefs/exoplanet_refs,/Users/sdaley/Work/Reading/References/MHDRefs/mhd_refs,/Users/sdaley/Work/Reading/References/CollidingWindsRefs/colliding_wind_refs,/Users/sdaley/Work/Reading/References/StellarWindRefs/stellar_wind_refs,/Users/sdaley/Work/Reading/References/SPMIRefs/SPMI_refs,/Users/sdaley/Work/Reading/References/OStarRefs/O_star_refs,/Users/sdaley/Work/Reading/References/MyRefs/my_refs}

\label{lastpage}

\end{document}